\documentclass[aps,twocolumn,superscriptaddress,amsmath,pra,10pt]{revtex4-2}

\usepackage{graphicx}
\usepackage[utf8]{inputenc}
\usepackage[colorlinks,allcolors=blue]{hyperref}
\usepackage{braket}
\usepackage{placeins}
\usepackage{amsmath}
\DeclareMathOperator{\Tr}{Tr}
\usepackage{xcolor}
\usepackage{comment}
\usepackage{soul}
\usepackage{soul}
\usepackage[normalem]{ulem}

\begin{document}

\newcommand{\mytitle}{Global Out of Time Order Correlators as a Signature of Scrambling Dynamics of Local Observables}

\title{\mytitle{}}

\author{Fabricio S. Lozano-Negro}

\affiliation{Instituto de Física Enrique Gaviola (CONICET-UNC)}

\affiliation{ Facultad de Matemática, Astronomía, Física y Computación, Universidad
Nacional de Córdoba, 5000, Córdoba, Argentina}

\author{Claudia~M.~Sánchez}

\affiliation{ Facultad de Matemática, Astronomía, Física y Computación, Universidad
Nacional de Córdoba, 5000, Córdoba, Argentina}

\author{Ana~K.~Chattah}

\affiliation{Instituto de Física Enrique Gaviola (CONICET-UNC)}

\affiliation{ Facultad de Matemática, Astronomía, Física y Computación, Universidad
Nacional de Córdoba, 5000, Córdoba, Argentina}

\author{Gonzalo~A.~Álvarez}

\affiliation{Centro Atómico Bariloche, CONICET, CNEA, S. C. de Bariloche, 8400,
Argentina}

\affiliation{Instituto de Nanociencia y Nanotecnologia, CNEA, CONICET, S. C. de
Bariloche, 8400, Argentina}

\affiliation{Instituto Balseiro, CNEA, Universidad Nacional de Cuyo, S. C. de
Bariloche, 8400, Argentina}

\author{Horacio~M.~Pastawski}

\affiliation{Instituto de Física Enrique Gaviola (CONICET-UNC)}

\affiliation{ Facultad de Matemática, Astronomía, Física y Computación, Universidad
Nacional de Córdoba, 5000, Córdoba, Argentina}
\begin{abstract}
Out-of-Time-Order Correlators (OTOCs) serve as a proxy for quantum information scrambling, which refers to the process where information stored locally disperses across the many-body degrees of freedom in a quantum system, rendering it inaccessible to local probes. Most experimental implementations of OTOCs to probe information scrambling rely on indirect measurements based on global observables, using techniques such as Loschmidt echoes and Multiple Quantum Coherences, via time reversal evolutions. In this article, we establish a direct connection between OTOCs with global and local observables in the context of NMR experiments, where the observable is the total magnetization of the system. We conduct a numerical analysis to quantify the differences in the evolution of both magnitudes, evaluating the excitation dynamics in spin ring systems with 8 to 16 spins, using a many-body Hamiltonian and long-range interactions. Our analysis decomposes the global echo into a sum of local echoes and cross-contributions, leading to local and global OTOCs. The results indicate that, after an initial transient period, local OTOCs determine the global ones. We observe that the difference between the average of local OTOCs and the global one, as well as their fluctuations, becomes negligible as the system size increases.   Thus, for large homogeneous systems, global and local OTOCs become equivalent. This behavior aligns with that observed in highly interacting or chaotic systems in several experiments.
\end{abstract}
\maketitle

\section{Introduction}

In recent years, the concept of Out-of-Time-Order Correlators (OTOC) captured the attention of many theoreticians and experimentalists alike ~\cite{GJaPaWi12,Sw18,LS+Rey19,Xu+Swi24} as a tool to identify manifestations of chaos in the scrambling of quantum information and the quantum butterfly effect~\cite{Shenker2014,BMEK17}. Quantum information scrambling refers to how local information propagates into many degrees of freedom, becoming encoded in complex correlations that prevents its recovery from local measurements. Such chaotic behavior is a crucial requirement for a quantum field theory to adequately describe the extreme classical instabilities induced by gravity in the proximity of a black hole~\cite{MalSSt16,Mald20}. 

Since OTOCs constitute a  quite broad category of mathematical objects, their physical significance and experimental relevance remained somewhat obscure until Alexei Kitaev~\cite{Ki17} realized that its basic concept was addressed in a paper by Larkin and Ovchinnikov~\cite{LaOv69}. They studied the effects of electron scattering in disordered superconductors,  observing that the same scattering processes that are responsible for the mean-free path also lead to the dynamical growth of the modulus square of a pair of initially commuting Heisenberg operators. Seeking a suitable quantum model with an extremely chaotic dynamics, Kitaev discarded the standard Heisenberg system in favor of Majorana Fermions with disorder and many-body infinite range interactions. This is now known as Sachdev-Ye-Kitaev (SYK) and should show an exponential growth of the OTO commutator. Nevertheless, an experimental approach to the problem seemed far-fetched, since it would involve the time-reverted dynamics of different many-body operators.

In a completely independent pathway, various forms of OTOCs were discovered and exploited by the nuclear magnetic resonance (NMR) community while not using explicitly this name. The backward evolution of Larmor precession of individual spins is the basis for the Hahn echo. There, the echo signal quantifies the failure to recover the initial state with the time scale $T_{2}$~\cite{Hh50}, which is due to the spreading of a local excitation through the spin-spin interactions. Decades later, the time reversal of the evolution driven by a many-body Hamiltonian enabled the observation of Magic Echoes and other generalized echoes~\cite{RPiWa70}. In particular, the initial quantum excitation scrambles into correlations between different spin projections, dubbed Multiple Quantum Coherences (MQC)~\cite{BMGPi85,MuPi86}.  The echoes that result from reverting the dynamics after imposing perturbative pulses of different strengths enable the extraction of a MQC  spectrum~\cite{C+CoRa05}. In this context, time-reversal protocols are regularly employed for spin counting, i.e. to determine the number of correlated spins.  
The protocol of polarization echoes enabled to address a set of individual spins \cite{ZMEr92,LUPa98}, and the Reduced Evolution of the Polarization Echoes sequence (REPE)~\cite{UPaLe98}, constituted a first attempt to scale down forward and backward dynamics with the purpose to disclose and quantify the role of the environment and perturbations which prevent the reversal of the many-body quantum dynamics within a time scale $T_3>T_2$. 

All of these protocols are now encompassed in the category of Loschmidt Echoes (LE)~\cite{JaPa01,GJaPaWi12}. Nowadays, LEs are one of the primary tools for studying quantum chaos, thermalization, excitation and information scrambling, as well as many-body localization. These studies are performed using both NMR and a variety of NMR-inspired innovative experimental techniques~\cite{ASu10,ASuK15,G+Rey17,YaNa18,Sw18,WeiChCa18,LS+BolRey19,Wei+Ca19,Sa+Pa20,NSCo20,YCZu20,Do-Al2021,Sw23}. Time reversal implementation also plays a key role in unmasking the environmental noise, eventually achieving its elimination through strategies broadly known as dynamical  decoupling~\cite{AlSu11,Ma+PaLu12,SoAlSu12,SuAl16,Sa+Ch16,G+Rey17}. 

As one experimental limitation, in particular in NMR measurements of solid state systems, resides in the difficulty of exciting and detecting individual spins, most of the experimental studies are based on global observables \cite{ASu10,ASuK15,G+Rey17,WeiChCa18,Wei+Ca19,Sa+Pa20,Do-Al2021,DoAl21,SaChPa22,SaPaCh23,ZwAl23}. Although local probing can be implemented in some very specific cases \cite{ZMEr92,LUPa98,LeChPa04,NSCo20,KuZwAl24}, this is not the general situation.  Thus, while a great majority of theoretical analyses address local observables on small systems, most LE/OTOC experiments still hinder their direct connection to the theoretical and analytical predictions. 


This article aims to verify that the LE/OTOCs that result from global observables, e.g. total polarization, evaluated over the entire system in these experimental implementations, accurately reflect the ensemble average of OTOCs that result from uncorrelated local observables, i.e. local spin projection. This equivalence hypothesis,  already stated in Ref.~\cite{Sa+Pa20}, is a crucial step to interpret the emergence of many-body irreversibility from the observation in terms of the dynamics of local excitations\cite{LUPa98,SaPaCh23}. Here, we demonstrate the validity of the equivalence hypothesis in the context of spin systems in NMR experiments in homogeneous lattices of equivalent spins, where the observable is the total magnetization, by employing a paradigmatic model for both numerical and analytical analyses. This confirmation is important because experimental implementations based on global control and readout are easier to perform compared to those requiring stringent and challenging conditions for local control and readout. Thus, already available experimental platforms based on global observables, such as NMR quantum simulations \cite{ASu10,ASuK15,WeiChCa18,Wei+Ca19,Sa+Pa20,Do-Al2021,DoAl21,SaChPa22,SaPaCh23,ZwAl23}, experiments with trapped ions \cite{G+Rey17,Le+Re19}, and ultra-cold polar molecules \cite{ZhSw23}, can be further exploited to probe information scrambling from local observables.

To pose a formal ground for the equivalence hypothesis, Section~\ref{OTOCnECHOES} first examines the analytical form of the specific OTOCs arising from MQC experiments and the spin-counting determined from the second moment of the MQC spectrum. There, we define the global and local observables, identifying in both cases, the particular contributions of OTOCs to local observables. Section \ref{NUMRESLT} presents the numerical evaluation of these magnitudes. Since it is impossible to solve a real system configuration, in which various dynamical regimes are present, we restrict the study to a paradigmatic model: a ring of spins with long-range interactions. This configuration partially mitigates the unavoidable finite-size effects. In a 1D system, long-range interactions are necessary to ensure coherence between multiple spin projection sub-spaces, while we introduce a local Zeeman disorder to crucially prevent symmetry-induced interferences. Given the small size of the systems that can be computationally studied, we show that the long-time behavior is representative of the equivalence between local and global observables, and thus is a more robust numerical metric when compared with the short and intermediate-time regimes, which are more sensitive to the system's particularities. We compare the time evolution and the equilibrium values of the global and local OTOCs for different system sizes, finding evidence that supports the validity of the mentioned equivalence as the complexity and size of the systems increase. This clarifies the role of long-range interactions and local disorder, emphasizing the long-time behavior. These results are discussed explicitly in Section\textbf{ }~\ref{Conclusions}, as they are of interest to a wide community pursuing related efforts~\cite{SchYao23,Sw23} in characterizing scrambling dynamics under different Hamiltonians in connection with existing experiments~\cite{Bra+Ol22,Xu+Swi24}.

\section{OTOCs and Echoes connection}

\label{OTOCnECHOES} The Out of Time Order (OTO) commutator is defined as, 
\begin{equation}
C_{\hat{V}\hat{W}}(t)=\Tr\left\{\left[\hat{W}(t),\hat{V}\right]^{\dagger}\left[\hat{W}(t),\hat{V}\right]\right\}.\label{otoc1}
\end{equation}
In the case of Hermitian Heisenberg operators $\hat{W}$ and $\hat{V}$ and unitary evolution, $\hat{W}(t)=e^{-i\hat{\mathcal{H}}t/\hbar}\hat{W}e^{i\hat{\mathcal{H}}t/\hbar}$ where $\hat{\mathcal{H}}$ is the system Hamiltonian, the expression can be rewritten in the form,
\begin{equation}
C_{\hat{V}\hat{W}}(t)=2\left(1-\Tr\left\{\hat{W}(t)^{\dagger}\hat{V}^{\dagger}\hat{W}(t)\hat{V}\right\}\right).\label{OTO}
\end{equation}
In the theoretical and numerical literature, $\hat{W}$ and $\hat{V}$ are generally considered local operators as, e.g. Pauli matrices, due to their direct interpretation as a measure of space-time propagation of quantum information~\cite{Garcia-Mata:2023}.

Considering that the operators $\hat{V}$ and $\hat{W}$ initially commute, the OTO commutator of Eq. (\ref{otoc1}) quantifies the degree by which the initially commuting operators fail to commute at time $t$ due to the scrambling of information induced by the Hamiltonian that drives the evolution. The correlator $F(t)$ defined as, 
\begin{equation}
F(t)=\Tr\left\{\hat{W}(t)^{\dagger}\hat{V}^{\dagger}\hat{W}(t)\hat{V}\right\},\label{otoc2}
\end{equation}
decays with time. According to Eqs.~\ref{OTO} and \ref{otoc2}, the correlator $F(t)$ and the commutator $C_{\hat{V}\hat{W}}(t)$ are related. In certain systems, a weak decay of $F(t)$ at short times determines the growth of $C_{\hat{V}\hat{W}}(t)$ with the same Lyapunov exponent that controls the corresponding classical system. Beyond the Ehrenfest time, however, the decay is dominated by Ruelle-Pollicott resonances~\cite{Pol15} and is stabilized by dephasing noise. As a result, both correlators serve as indicators of information scrambling and quantum chaos \cite{RGGa17,Gm+Wi18,FGmJaWi19}. Nonetheless, assessing chaos lies beyond the scope of this paper, as we focus on the equivalence between OTOCs defined through local and global observables under general statistical considerations.

\begin{figure}[t]
\centering \includegraphics[width=1\columnwidth]{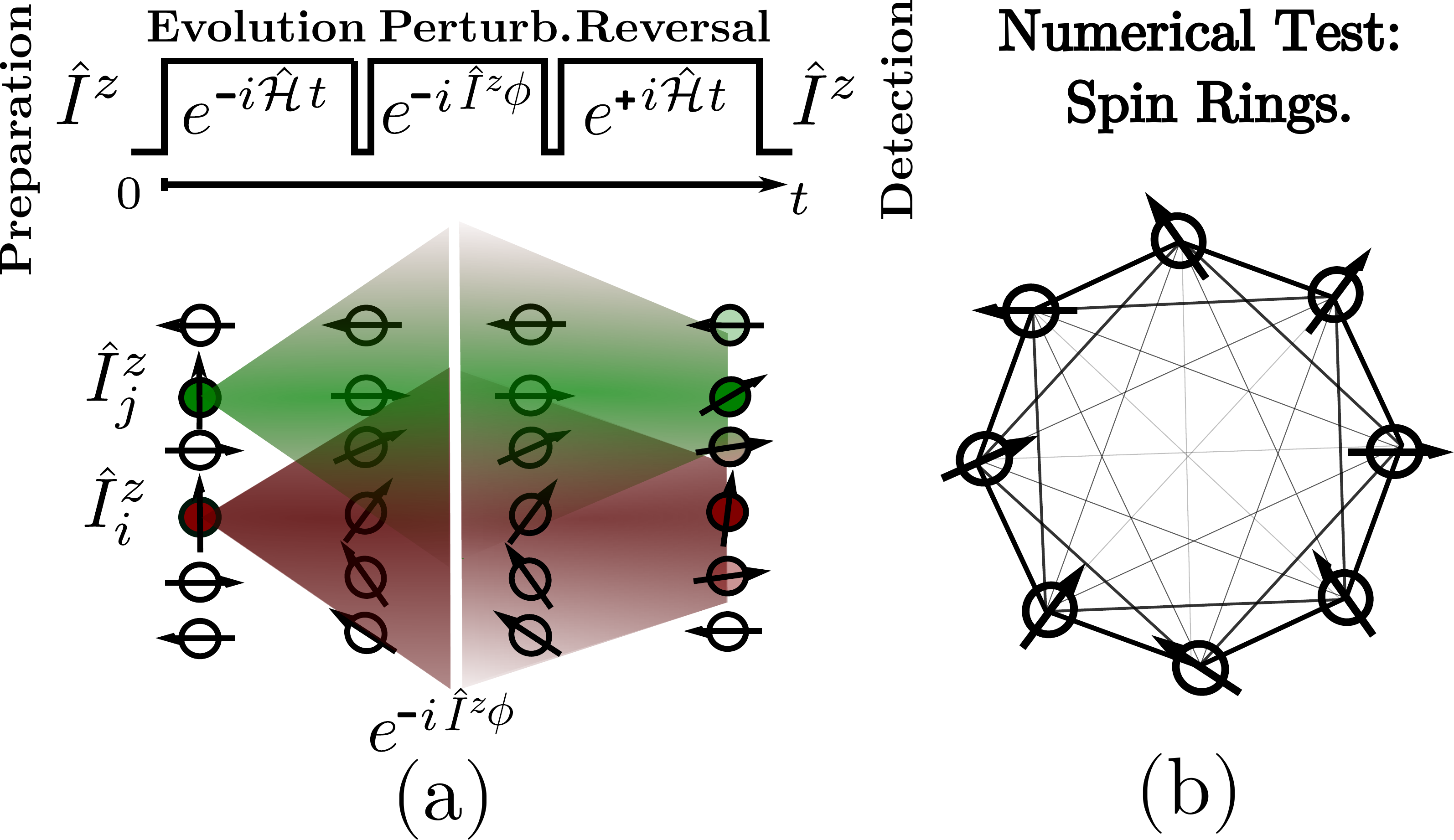} \caption{(a) Top: Evolution sequence of a MQC experiment. Bottom: Pictorial representation of the equivalence hypothesis regarding the global Loschmidt echo and MQC experiment. This picture shows the complete evolution of two contributions to the total magnetization, expressly at site $i$ and $j$ in green and maroon colors. After the evolution, perturbation, and reversal protocol, each magnetization return mainly to the original sites. The magnetization that arrives at different sites cancel each-other (sites with a superposition of green and maroon). The global magnetization $\hat{I}^{z}$ measured at the time reversal echo is given by all the local echoes, as the polarization coming from other sites interfere destructively. (b) Scheme of the considered spin ring systems for the numerical simulations. The range of the inter spin couplings are chosen in terms of the ``bond-distance'' between spins, $\propto\frac{1}{r^{\alpha}}$ for $\alpha=1,2,3$. }
\label{fig:01} 
\end{figure}

The correlator $F(t)$ involves a time-reversal protocol and its calculation can be thought of as an experiment in which $\hat{V}$ sets a quantum excitation that evolves for a time $t$, and then it suffers the action of $\hat{W}$. Later, it follows a time reversal evolution before a measurement is applied ($\hat{V}^{\dagger}$) \cite{Sa+Pa20,LoZgPa21,Do-Al2021}. Under this view, $F(t)$ has the form of a Loschmidt echo with a brief perturbation $\hat{W}=\exp[-i\hat{\Theta}\Delta t]$, where $\hat{\Theta}$ is a Hamiltonian (e.g. describing a global Zeeman field~\cite{BMGPi85,MuPi86} or a field gradient~\cite{LoZgPa21}) that acts for a very short period $\Delta t$ (pulse labeling) after the forward evolution lapse \cite{Sa+Pa20}.
The operator $\hat{W}(t)$ can be interpreted as a Loschmidt echo evolution operator as it includes a forward evolution, a perturbation, and a backward evolution. Consequently, $\hat{W}(t)\hat{V}\hat{W}^\dagger(t)$ is the Loschmidt echo dynamics of the operator $\hat{V}$.

 In actual NMR experiments on many-body systems, there is also a small but mostly uncontrollable perturbation, $\hat{\Sigma}$, that persists during the whole time reversal period. Thus, there is no possible factorization in the above form \cite{Sa+Pa20}. As a consequence, the recovered signal when $\hat{W}=\hat{\mathcal{I}}$  has an overall decay with a time-scale $T_{3}$ defined as the time at which $F(t)$ is half of its initial value, an example is shown in Appendix~\ref{sm:expres}. In single-particle semi-classical models that idealize this situation, the decay is exponential. For weak  $\hat{\Sigma}$ the decay rate depends on the perturbation strength according to the Fermi golden rule, as expected. However, once $1/T_{3}$ reaches the classical Lyapunov time-scale \cite{Pa+00,JaPa01}, it remains constant for a wide range of $\hat{\Sigma}$, i.e. it becomes perturbation independent. These result contrasts with the observation of many-spin systems. When reversible many-body interactions dominate the dynamics $T_{3}$ depends on the Hamiltonian and the Fermi golden rule regime is never reached,  i.e. $T_3$ becomes perturbation-independent and precisely proportional to spin-spin characteristic time $T_{2}$, defined through the inverse second moment of the Hamiltonian driving the dynamics \cite{Sa+Pa20}. The stability of $T_3$ towards decoherent noise has a resemblance to the Ruelle-Pollicott regime of $F(t)$~\cite{Gm+Wi18}. Besides its conceptual value, the global Loschmidt echo magnitude is widely used in experimental setups as a practical tool for the normalization of the signal for quantifying OTOCs and for characterizing the many-body dynamics and its information scrambling \cite{Do-Al2021,DoAl21,SaChPa22,SaPaCh23}.

The dynamics of OTOCs as a measure of growth in ``size'' and complexity of the spreading of an initial local operator have been studied in closed and open systems~\cite{SchYao23}, linking this complexity with the system's sensitivity to decoherence \cite{Do-Al2021,DoAl21}, and a manifestation of quantum chaos \cite{GaJaWi22}. The dynamical regimes for the OTOCs can be separated into short, intermediate, and long times. The short and intermediate times are highly dependent on the Hamiltonian and on the particular nature of the initial operators (local or global). At long times, the OTOCs of a finite system oscillate or, for highly chaotic systems, fluctuate around a mean value~\cite{FGmJaWi19}.

In Ref.~\cite{Sa+Pa20} we proposed that the information extracted from global OTOCs is indicative of the behavior of the local observables. Specifically, we use the Lanczos expansion of the density matrix dynamics to infer the intermediate time behavior of local OTOCs from the global observables. The main hypothesis was that the latter is mainly composed of a set of almost identical local magnitudes plus small interference terms that tend to cancel out. Then, Zhou and Swingle studied the contribution of local OTOCs to global observables, arguing that, in an expansion, the ``diagonal'' terms are those that contribute the most \cite{Sw23}. As with most numerical results, their verification was restricted to the dynamics of 1D chains of spins. In the present work, we take a different perspective on this matter, making a particular focus on the experimental observables in NMR experiments, which are the average magnetization associated with generalized time reversal echoes.

\begin{center}
\begin{figure*}[t]
\centering \includegraphics[width=1\textwidth]{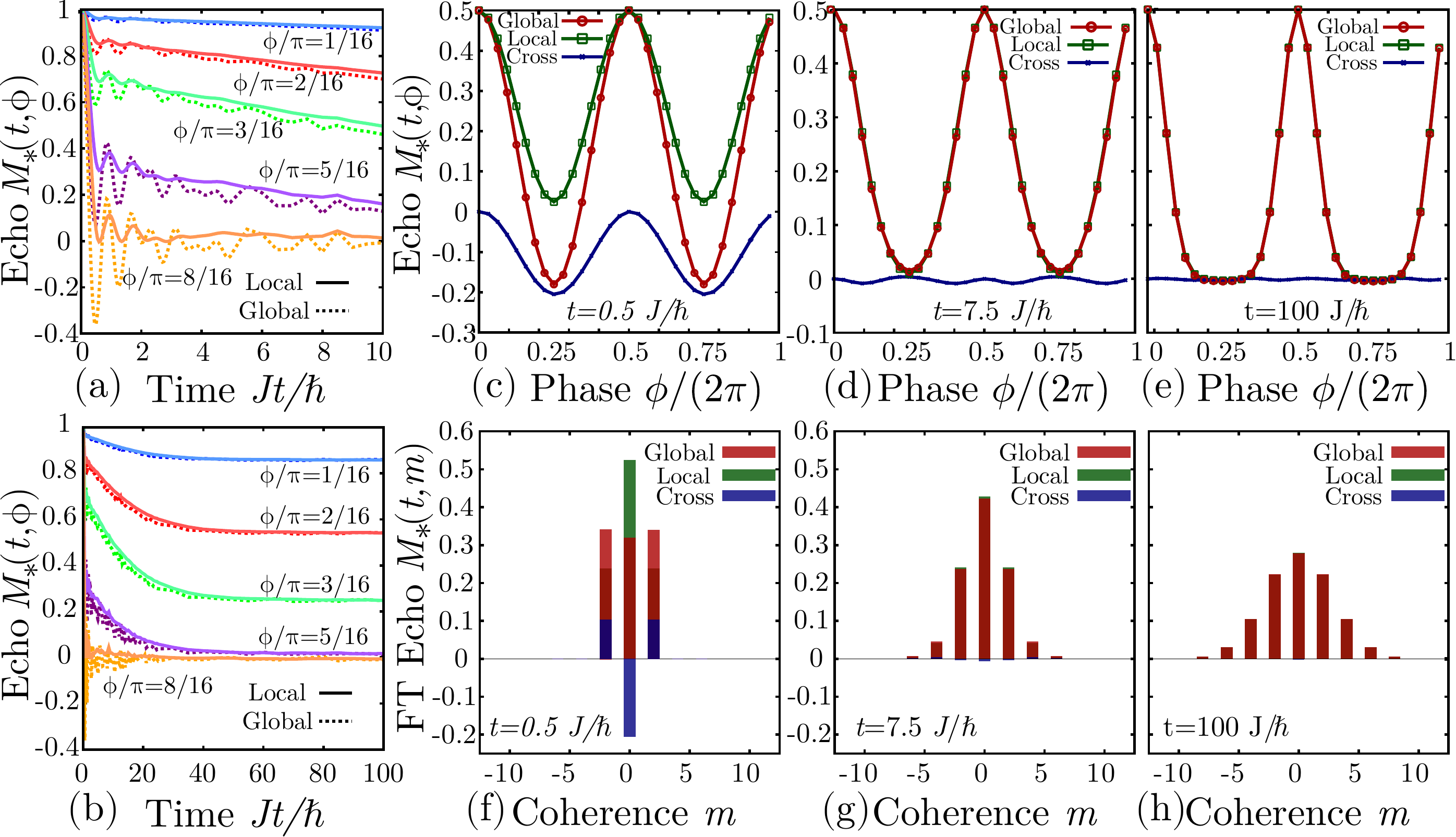}
\caption{(a-b) Global echoes $M_{G}(t,\phi)$ (dashed curves) and local ones $M_{L}(t,\phi)$ (solid curves), as a function of time for five representative values of $\phi=\{\pi/16,2\pi/16,3\pi/16,5\pi/16,8\pi/16\}$ (colors). Panel (a) shows the short and intermediate time behavior of the echoes, while panel (b) shows the complete decay. (c-e) $M_{G}(t,\phi)$ (green), $M_{L}(t,\phi)$ (red) and $M_{CT}(t,\phi)$ (blue) as a function $\phi$ for fixed times. (f-h): Distributions $M_{G}(t,m)$ (red), $M_{L}(t,m)$ (green) and $M_{CT}(t,m)$ (blue) obtained from the FT of curves in (c-e). The considered times are (c/f) $t=0.5J/\hbar$, (d/g) $t=7.5J/\hbar$, (e/h) $t=100J/\hbar$. All panels are computed using a spin ring of $N=16$ and $\alpha=3$.}
\label{fig:NumImp} 
\end{figure*}
\par\end{center}

\begin{center}
\par\end{center}

\subsection{Generalized Echoes and MQC in NMR}

In numerous real situations, particularly those connected to many-body spin systems analyzed through the solid-state NMR techniques, the acquired signal is related to global operators. In NMR, the observable is the total magnetization of the sample, which is proportional to the total spin magnetization denoted as $\hat{I}^{z}=\sum\hat{I}_{i}^{z}$, which adds the contribution of individual spin polarizations. The direction $z$ is determined by the external magnetic field. Correspondingly, the initial condition is usually the equilibrium magnetization of an ensemble of polarized spins, in which the density matrix is $\hat{\rho}(t=0)\propto\hat{I}^{z}$, i.e. also a global operator describes the initial excitation. Notice that $\hat{\rho}(t=0)$ is determined by the Boltzmann equilibrium in the high temperature limit, where terms proportional to the identity do not contribute to an observable signal and are therefore omitted \cite{ASu10,ASuK15,WeiChCa18,Wei+Ca19,Sa+Pa20,Do-Al2021,DoAl21,SaChPa22,SaPaCh23,ZwAl23}.

In a solid sample, the natural interaction between spins $I=1/2$ is given by the dipolar Hamiltonian \cite{Sl90}. A huge number of protocols of radio frequency pulse sequences were developed to engineer the spin-spin Hamiltonian for practical applications. 
Typically, a protocol design is based on the average Hamiltonian theory, which is typically based on the Magnus expansion and/or the Floquet approximation, as referenced in several works \cite{Hb76,Er87,MuPi87b}. The key feature of these approaches is the ability to reverse the quantum dynamics by implementing a change in the sign of the acting Hamiltonian.



The observation of MQC (or generalized echoes) for spin counting purposes, involves three time periods, see the upper panel of Fig. \ref{fig:01}a. An initial excitation evolves (forward) with a specifically tailored Hamiltonian $\hat{\mathcal{H}}$; this is followed by a brief encoding period that serves to imprint a different phase to different spin-projection sub-spaces (phase labeling of the quantum coherences). Finally, a time reversal is achieved by imposing a dynamics with $-\hat{\mathcal{H}}$ Hamiltonian. Experimentally, the final observable and the initial state are proportional to the total magnetization $\hat{I}^{z}$ operator (i.e. $\hat{V}\propto\hat{I}^z$). The phase encoding performed through a rotation around the $z$ axis plays the role of the perturbation ($\hat{\Theta}\propto\hat{I}^z$ and therefore $\hat{W}=e^{-i\phi\hat{I}^z}$) in the OTOCs protocol. 
According to the ``pseudo-pure state'' description~\cite{CoFH97,Co+00}, this initial thermal state $\hat{\rho}(t=0)\propto\hat{I}^{z}=\sum\hat{I}_{i}^{z}$ can be thought of as a sum of individual magnetized spins with no correlations among them \cite{ASu10,ASuK15,Sa+Pa20,Do-Al2021,DoAl21,SaChPa22,SaPaCh23,ZwAl23}. Each interacting spin will undergo a collective evolution and, after the perturbation and the backward dynamics, the resulting collective state contributes with magnetization not only at its original site but also to neighboring spins. Our primary hypothesis is that the main contribution to the global echo (total magnetization) arises from the individual magnetization of each spin returning to its original site. This concept is illustrated schematically in the Fig. \ref{fig:01}a. We conjecture that any magnetization not returning to the original spin site will cancel each other out, as they arrive with ``random'' phases.

The generalized echo sequence of Fig. \ref{fig:01} produces a global observable signal denoted as $M_{G}$, which is measured after a final read-out pulse (not appearing in the figure). This echo can be summarized in the following equation, 
\begin{eqnarray}
M_{G}(t,\phi)=\frac{1}{\Tr\{\hat{I}^{z}\hat{I}^{z}\}}\Tr\{\hat{I}^{z}(t)\hat{R}^{\dagger}\hat{I}^{z}(t)\hat{R}\}\label{echo}
\end{eqnarray}
where $\hat{R}=e^{-i\phi\hat{I}^{z}}$, $\hat{I}^{z}(t)=e^{-i\hat{\mathcal{H}}t}\hat{I}^{z}e^{i\hat{\mathcal{H}}t}$,
and the normalization $\Tr\{\hat{I}^{z}\hat{I}^{z}\}=N2^{N-2}$ ensures $M_{G}(0,\phi)=1$. Here, the operators $\hat{R}$ and $\hat{I}^{z}(0)$ commute, however, this is no longer true once the state $\hat{I}^{z}(0)$ evolved into $\hat{I}^{z}(t)$. This led to the concept of the magnetization scrambling. In the experiments, the phase $\phi$ is varied between $0$ and $2\pi$ in $2^{M}>m_{max}$ steps, enabling the acquisition of the multiple quantum coherence distribution $M_{G}(t,m)$ by Fourier transforming the signal $M_{G}(t,\phi)$ as a function of $\phi$, with $m$ ranging from $-m_{max}$ to $m_{max}$ (see Appendix~\ref{appA}). This distribution reflects the superposition of states in different total magnetization subspaces.

In the following section, we clarify the connection of Eq. \eqref{echo} with a global OTOC and rewrite it as a combination of a set of local OTOCs. 

\begin{center}
\begin{figure*}
\includegraphics[width=1\textwidth]{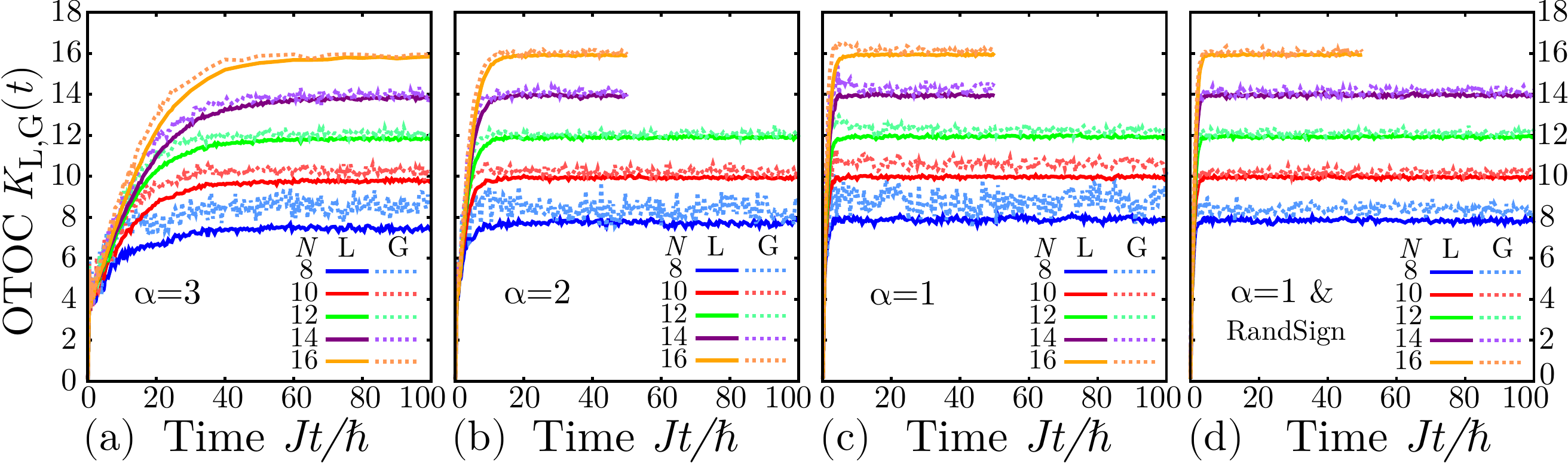}
\caption{ Time evolution of the local OTOC $K_{L}(t)$ (solid curves), and the global OTOC $K_{G}(t)$ (dashed curves), for a ring-system with interactions given by Eq. \eqref{eq:Ham}. Interactions are of the form, (a) $D_{ij}\propto\frac{J}{|r_{ij}|^{3}}$, (b) $D_{ij}\propto\frac{J}{|r_{ij}|^{2}}$, (c) $D_{ij}\propto\frac{J}{|r_{ij}|}$, (d) $D_{ij}\propto\frac{\pm J}{|r_{ij}|}$ with random signs. }
\label{fig:02} 
\end{figure*}
\par\end{center}

\begin{figure}
\centering \includegraphics[width=1\columnwidth]{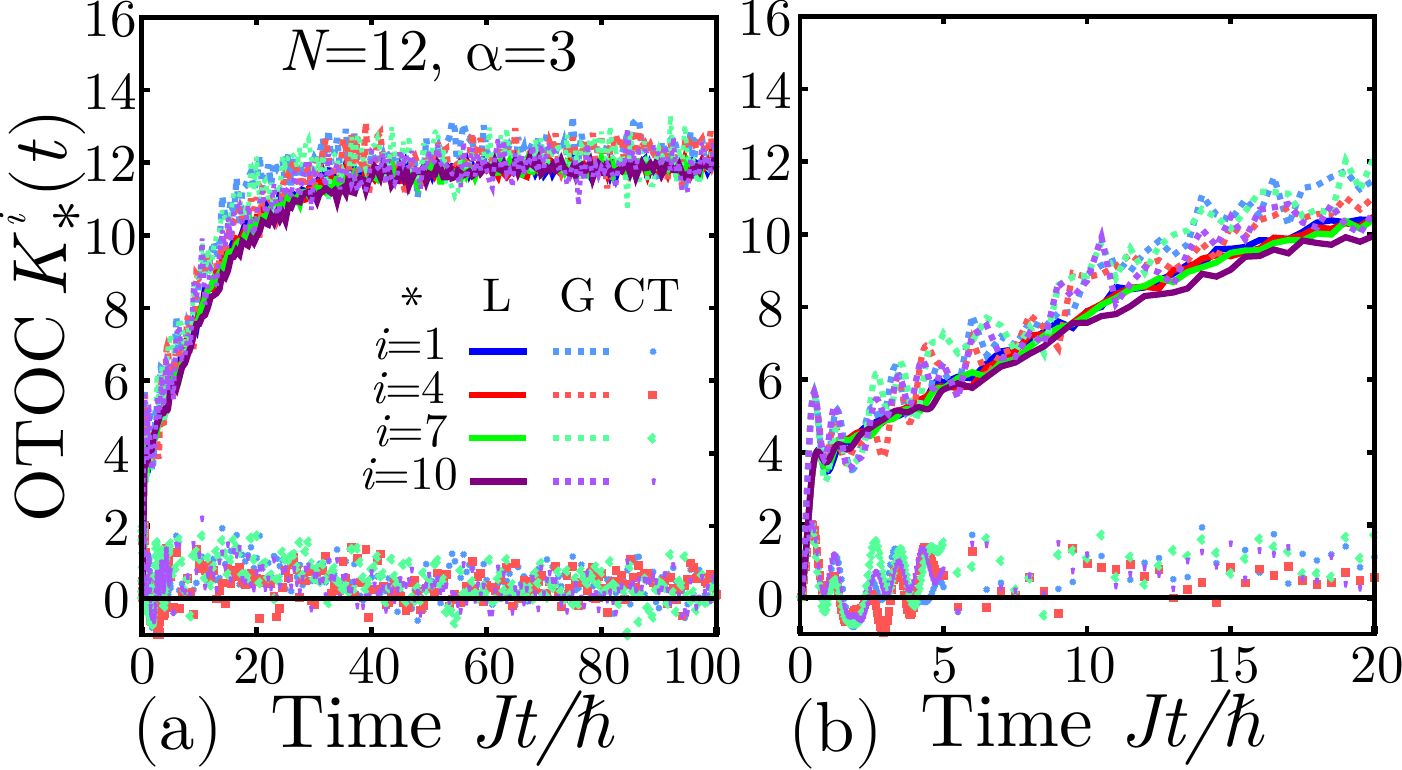}
\caption{Time evolution of individual realizations per site $K_{*}^{i}$. The supra-index $i$ represents the site (colors), while the asterisk $*$ indicates whether the OTOC is Global (G, dashed curves), Local (L, solid curves), or Cross Terms (CT, symbols). Panel (b) shows a short-time zoom of the figure on the left. All data corresponds to $D_{ij}=J/|r_{i,j}|^{3}$ and $N=12$.}
\label{fig:02a} 
\end{figure}

\subsection{Local and Global observables}

The echo in Eq. \eqref{echo} can be expressed in terms of local echoes,
\begin{eqnarray}
M_{G}(t,\phi) & = & \frac{1}{N2^{N-2}}\Tr\{\hat{I}^{z}(t)\hat{R}^{\dagger}\hat{I}^{z}(t)\hat{R}\}\label{MGLyC}\\
 & = & \frac{1}{N2^{N-2}}\sum_{i,j}\Tr\{\hat{I}_{i}^{z}(t)\hat{R}^{\dagger}\hat{I}_{j}^{z}(t)\hat{R}\}\nonumber \\
 & = & \frac{1}{N2^{N-2}}\sum_{\substack{i}
}\Tr\{\hat{I}_{i}^{z}(t)\hat{R}^{\dagger}\hat{I}_{i}^{z}(t)\hat{R}\}\nonumber \\
 & + & \sum_{\substack{j\neq i}
}\Tr\{\hat{I}_{i}^{z}(t)\hat{R}^{\dagger}\hat{I}_{j}^{z}(t)\hat{R}\},\nonumber
\end{eqnarray}
where each local component of the polarization emerges from a local echo (initial excitation at the same site) plus cross contribution terms. This allows the separation of the global echo into two terms containing local echoes and cross terms
\begin{eqnarray}
M_{G}(t,\phi) & = & \sum_{i}\left(M_{L}^{i}(t,\phi)+M_{CT}^{i}(t,\phi)\right)\\
 & = & M_{L}(t,\phi)+M_{CT}(t,\phi).
\end{eqnarray}

By applying the Fourier transformation of $M_{G}(t,\phi)$ with respect to $\phi$, the multiple quantum coherence (MQC) distribution $M_{G}(t,m)$ is obtained. The global MQC distribution can also be written in terms of contributions from local and cross-terms, $M_{G}(t,m)=M_{L}(t,m)+M_{CT}(t,m)$. From the early NMR literature \cite{BMGPi85,MuPi86,Munowitz} and subsequent extensive usage, it is established that the second moment of this distribution is a statistical measure of the number of correlated spins, here denoted as $K_{G}$ \cite{ASu10,ASuK15,Sa+Pa20,Do-Al2021,DoAl21,SaChPa22,SaPaCh23,ZwAl23}. The global number of correlated spins (sometimes called cluster size)
$K_{G}$, is expressed as
\begin{eqnarray}
K_{G}(t) & = & 2\sum_{m}m^{2}M_{G}(t,m)\label{KGLyC}\\
 & = & 2\sum_{m}m^{2}(M_{L}(t,m)+M_{CT}(t,m))\nonumber \\
 & = & K_{L}(t)+K_{CT}(t),\nonumber 
\end{eqnarray}
from which, the contributions from local echoes ($L$ subindex) and cross-terms ($CT$ subindex) are explicitly discriminated.

The global cluster size can also be expressed as the OTOC \cite{Kh97,Do-Al2021,DoAl21},
\begin{equation}
K_{G}(t)=-\frac{2}{N2^{N-2}}\Tr\left\{\left[\hat{I}^{z},\hat{I}^{z}(t)\right]\left[\hat{I}^{z},\hat{I}^{z}(t)\right]\right\},\label{eq:KGOTOC}
\end{equation}
as well as the corresponding local and cross-terms contribution, 
\small
\begin{eqnarray}
K_{L}(t) &=& \frac{-2}{N2^{N-2}}\left(\sum_{i,k}\Tr\left\{\left[\hat{I}_{k}^{z},\hat{I}_{i}^{z}(t)\right]^{2}\right\}\right.\nonumber \\
 &  & \left.+\sum_{\substack{i,q,k\\
q\neq k
}
}\Tr\left\{\left[\hat{I}_{q}^{z},\hat{I}_{i}^{z}(t)\right]\left[\hat{I}_{k}^{z},\hat{I}_{i}^{z}(t)\right]\right\}\right)\label{eq:KLOTOC}\\
K_{CT}(t) &=& \frac{-2}{N2^{N-2}}\sum_{\substack{i,j,k,q\\ i\neq j}
}\Tr\left\{\left[\hat{I}_{q}^{z},\hat{I}_{i}^{z}(t)\right]\left[\hat{I}_{k}^{z},\hat{I}_{j}^{z}(t)\right]\right\}.\label{eq:KCOTOC}
\end{eqnarray}
\normalsize
A detailed derivation of these expressions is provided in the appendix \ref{appA} and \ref{appB}. An interesting feature to note is that, in a similar fashion as the global echo can be thought as the sum over different initial conditions $\hat{I}_{i}^{z}$, the equivalent procedure can be done with both $K_{L}(t)$ and $K_{CT}(t)$. Then, the sum over sites $i$ can be separated in the previous expressions, defining the on-site averages, 
\begin{eqnarray*}
K_{G}(t) & = & \sum_{i}K_{G}^{i}(t)/N,\\
K_{L}(t) & = & \sum_{i}K_{L}^{i}(t)/N,\\
K_{CT}(t) & = & \sum_{i}K_{CT}^{i}(t)/N.
\end{eqnarray*}
Notice that while the local OTOCs corresponding to a site $i$, $K_{L}^{i}(t)$ are composed for both, the (so-called in ~\cite{Sw23}) diagonal terms $\Tr\left\{\left[\hat{I}_{k}^{z},\hat{I}_{i}^{z}(t)\right]^{2}\right\}$ and non-diagonal terms $\Tr\left\{\left[\hat{I}_{q}^{z},\hat{I}_{i}^{z}(t)\right]\left[\hat{I}_{k}^{z},\hat{I}_{i}^{z}(t)\right]\right\}$, the cross-contribution corresponding to site $i$, $K_{CT}^{i}(t)$, only have non-diagonal terms. Consequently, the same property is valid for their on-site averages $K_{L}(t)$ and $K_{CT}(t)$.

Numerically, we can compute the $i$-contribution of these magnitudes $K_{L}^{i}(t),$ $K_{G}^{i}(t)$, $K_{CT}^{i}(t)$ using the echo sequence, shown in Fig. \ref{fig:01}, for an initial state $\hat{I}_{i}^{z}$. Therefore, starting from an excitation localized at site $i$, and observing the magnetization evolution and the subsequent return (Loschmidt echoes) to every site $j$, we can reconstruct $M_{L}^{i}(t,\phi)$ and $M_{CT}^{i}(t,\phi)$, enabling us to compute separately the local and cross-terms OTOCs, and the addition of both, which is the global OTOC.

\begin{figure}
\centering \includegraphics[width=1\columnwidth]{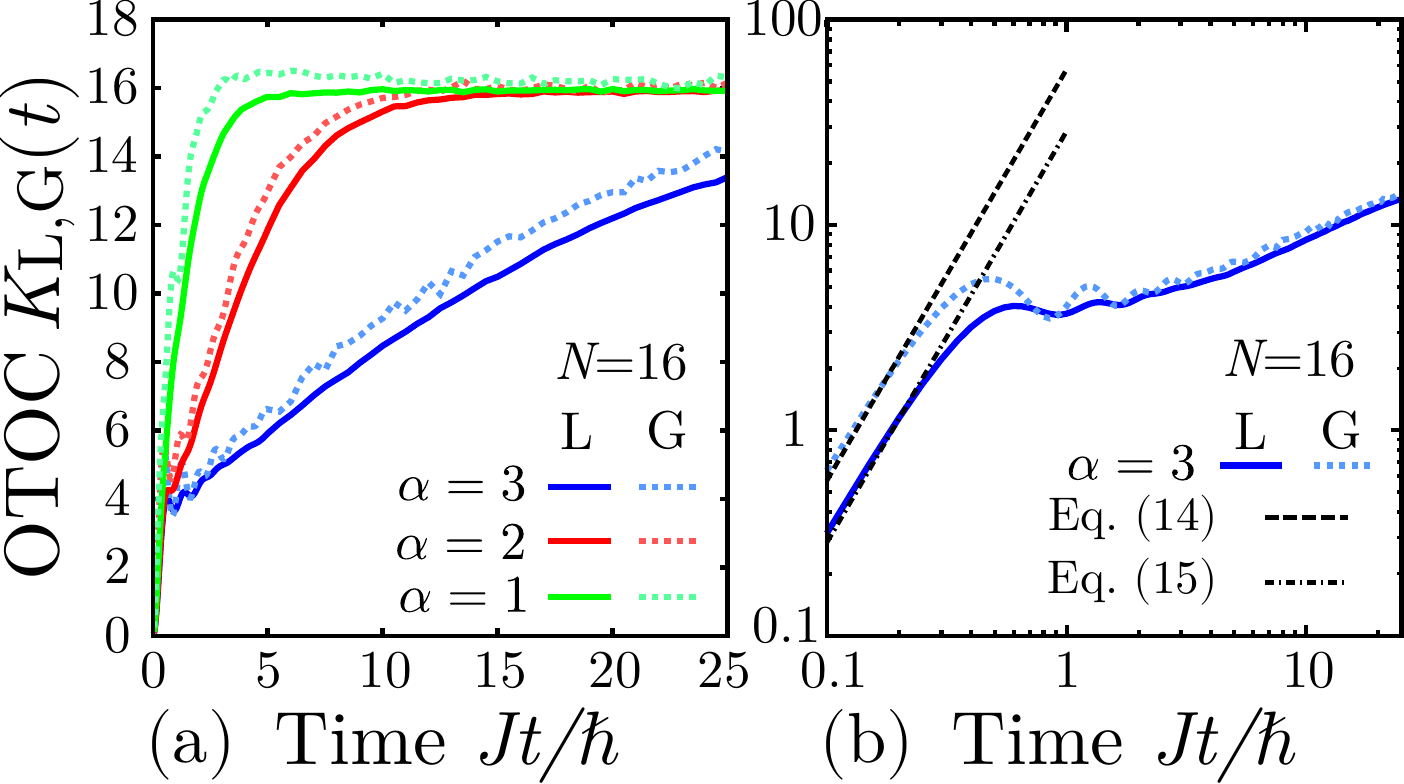}
\caption{(a) Evolution of $K_{L}(t)$ (solid curves) and $K_{G}(t)$ (dashed curves) for times prior to saturation. Interactions of the form $D_{ij}\propto1/|r_{ij}|^{\alpha}$ with $\alpha=\{1,2,3\}$, and a ring with $N=16$. (b) Zoom at the short time behavior for $\alpha=3$ in log-log scale. In black dashed lines the analytical expression given by Eq.~\eqref{eq:KG-shorttime} and~\eqref{eq:KL-shorttime} are shown. }
\label{fig:02b} 
\end{figure}

\section{Numerical results}

\label{NUMRESLT}

Global, local, and cross-term magnitudes described in the previous section were calculated considering the paradigmatic model system shown in Fig. \ref{fig:01}b: a ring of $N$ spins $1/2$ interacting through a long-range Double Quantum Hamiltonian, 
\begin{equation}
\hat{\mathcal{H}}=\sum_{i}h_{i}\hat{I}_{i}^{z}+\sum_{\substack{i,j\\
i<j
}
}D_{ij}\left[\hat{I}_{i}^{x}\hat{I}_{j}^{x}-\hat{I}_{i}^{y}\hat{I}_{j}^{y}\right].\label{eq:Ham}
\end{equation}
This double quantum Hamiltonian is experimentally engineered using NMR pulse sequences developed in the early references \cite{BPi86,MuPi87b}, and further modified for scaled interactions as described in \cite{SaPaCh23}.

Extending further the modeling qualities, we assume interactions between spins $D_{ij}$ with different dependencies $D_{ij}=J/|r_{ij}|^{\alpha}$ on the ``bond-distances'' $r_{ij}$ for $\alpha=1,2,3$, where $\alpha=3$ is the usual dipolar case. It is important to note that $r_{ij}$ is defined as the minimum number of sites between the two spins rather than as a geometric distance. This definition is pivotal for preserving system homogeneity across varying values of $N$. Since we mostly use non-random $D_{ij}$'s, it is crucial to introduce random fields $h_{i}$ to break the high symmetry of the ring and wash away recurrences.
We also choose $D_{ij}$'s ranging between $[-J/2,J/2]$. The interactions in typical molecular Hamiltonians have a sign that depends on the angle between the bond and the external field. However, for this study, we adopt a uniform sign convention, as would be the case in a Ferrocene ring~\cite{LUPa98,Sa+Ch16}. Additionally, we explore the incorporation of random sign assignments in the coupling, considering $\alpha=1$ as a paradigmatic case.

The initial local excitation has the form $\hat{\rho}_{0}\propto\hat{I}_{i}^{z}$, and evolves, as in a typical MQC sequence (Fig.~\ref{fig:01}a) with the Hamiltonian defined in Eq.~\eqref{eq:Ham}. Since a form of self-average is naturally present in a global observable, a single disorder realization is considered. The evolution was done following the Trotter-Susuki~\cite{Su76,DRae83,DRaeM04} and quantum parallelism algorithm~\cite{ADLP08,DBZgPa13}. As pointed out above, by repeating the simulations for all the possible initial sites $i$, the global and local Loschmidt echoes can be computed, from which the OTOCs $K_{*}$ are derived, where {*} represents ({G,L,CT})
.

Fig. \ref{fig:NumImp}(a-b) shows the echoes $M_{G}(t,\phi)$ (dashed curves) obtained by performing the numerical evolution for a system of $N=16$ spins with $\alpha=3$, starting from the initial condition (i.e. the excitation) at the different sites, $i$, and adding all the signals regardless at which site it is detected. This global magnitude represents the experimental observable given by the total magnetization $I^z$, Eq. \eqref{echo}, see Appendix~\ref{sm:expres}. Together with the global echo, Fig. \ref{fig:NumImp} (a and b) also displays local echo $M_{L}(t,\phi)$, which is only accessible numerically, for different perturbations (phases), as a function of time. For short times, differences between local and global echoes are still noticeable, but these become smaller as the system evolves, becoming indistinguishable at long times. Appendix \ref{sm:EchoAway} presents a more detailed analysis of the echoes to different sites. 

When analyzing the recovered signal for a fixed time as a function of phases, Figs. \ref{fig:NumImp}(c-e), it becomes clear that at short times, (c), the differences between global and local are still appreciable, but these decrease over time. This is evidenced by the cross terms, which are practically zero in (e). This temporal behavior is also reflected in the contributions of the different terms, Global, G; Local, L; and Cross-Terms, CT, to the corresponding distribution of coherence $M_{*}(t,m)$, seen in Fig. \ref{fig:NumImp} (f-h). The second moments of these distributions are proportional to $K_{G}(t)$, $K_{L}(t)$ and $K_{CT}(t)$.

Fig. \ref{fig:02} shows the time evolution of the number of correlated spins, global or local $K_{G}(t)$ and $K_{L}(t)$ (dashed and solid lines respectively) for different sizes of the ring $N=8-16$. These are obtained by averaging over the individual realizations at different sites, $K_{*}^{i}$. The evolution of individual $K_{*}^{i}$'s is exemplified in Fig. \ref{fig:02a} for the case $D_{ij}=J/|r_{i,j}|^{3}$ and $N=12$. $K_{*}^{i}$ present the same behavior of the total values, but differing in fluctuations. From left to right, Fig. \ref{fig:02} displays the results for all values of $\alpha$ from 3 to 1, plus $\alpha=1$ and random signs. 
We observe that, after the initial regime, curves $K_{L}$ and $K_{G}$ representing the growth in the number of correlated spins, differ by less than $10\%$. We use the long-time saturation value and its fluctuations to quantify this difference as the number of spins in the system $N$ increases. As $\alpha$ decreases, the magnitudes $K_{*}$ reach the saturation values at shorter times, due to the stronger couplings. Typically, the saturation times $t_{s}$ are $Jt_{s}/\hbar\approx50$ for interaction $\propto1/r^{3}$, $Jt_{s}/\hbar\approx20$ for $\propto1/r^{2}$ and $Jt_{s}/\hbar\approx10$ for $\propto1/r$, a more detailed analysis would yield $t_{s}$ depending on $N$ and $\alpha$. 
We can observe that both $K_{G}(t)$ and $K_{L}(t)$, at saturation times, tend towards a value around the system size $N$.

In the limit of large $\alpha$ the interaction among nearest neighbors predominates, leading to a chain-like behavior. In this limit, the Double Quantum Hamiltonian exclusively generates only second-order coherences \cite{Feld97,RfSaPa09,CaRaCo07,CaRaCo07b}, 
and the difference between global and local OTOCs should be more relevant. Conversely, in the limit of very small $\alpha$, the interaction extends infinitely, and in the case of couplings with random signs, it should behave like the SYK model~\cite{Ki17}. The fact that for larger systems the cross terms become relatively less important means that adding pathways to the dynamics increases the destructive interferences of cross-terms. This suggests, that one could enhance these destructive interferences by allowing random signs in $D_{ij}$, as is exemplified in Fig. \ref{fig:02}(d). Indeed, pseudo-random signs appear in an actual crystal due to the different directions of the coupling.

At short times and intermediate times, the growth of the Local and Global OTOCs looks slightly different, as depicted in Fig. \ref{fig:02b}. This fact can be traced back to the particular interference patterns in the spin dynamics of the DQ Hamiltonian during time reversal. Components failing to return to their original sites within this brief time-frame exhibit a strong tendency to return to their adjacent sites, with specific phase relationships. 
Mathematically, the very short time difference can be analyzed using the Baker-Campbell-Hausdorff expansion \cite{Er87}. After performing some algebraic manipulation, it can be shown (as elaborated in Appendix \ref{SM:ShortTime}) that for the DQ Hamiltonian, both the global and local OTOCs exhibit quadratic behavior at short times, with coefficients differing only by a factor of two: 
\begin{eqnarray}
K_{G}(t) & \approx & \frac{32}{N}t^{2}\hbar^{2}\sum_{i,j}D_{ij}^{2},\label{eq:KG-shorttime}\\
K_{L}(t) & \approx & \frac{16}{N}t^{2}\hbar^{2}\sum_{i,j}D_{ij}^{2}.\label{eq:KL-shorttime}
\end{eqnarray}
Consequently, $K_{L}(t)\approx K_{CT}(t)$ at short times. These expressions align well with the numerical findings, as shown in Fig. \ref{fig:02b}(b). It is also seen that growth of the OTOCs accelerates when the exponent $\alpha$ becomes smaller, as it increases the value of $\sum_{i,j}D_{ij}^{2}$. Moreover, this behavior should not change when random signs are included in the values of $D_{ij}$, as seen in Fig. \ref{fig:02}(d).

At intermediate times, after this initial quadratic expansion, the complexity of the Hamiltonian makes itself evident and the growth law changes depending on the exponent $\alpha,$ a behavior that it is theoretically expected~\cite{Zh+Sw20,HaFi14,ShPar16,Foss+Gor15}. In Fig. \ref{fig:02b} we observe that this growth law for local and global OTOCs shows the same behavior, making their relative difference $K_{CT}(t)/K_{G}(t)$ rapidly smaller as time increases. However, the relatively small growth window makes it difficult to assign a particular law, leaving only the saturated regime to systematically study the dependence of the cross terms $K_{CT}(t)/K_{G}(t)$ with $N$. Indeed, the clear intermediate time dynamics observed in the experiments are a consequence of the exponential increase of the number of states of the Hilbert space enabled by the dynamics in 3D crystals \cite{Sa+Ch16,Sa+Pa20}.

\begin{center}
\begin{figure*}[t]
\centering \includegraphics[width=1\textwidth]{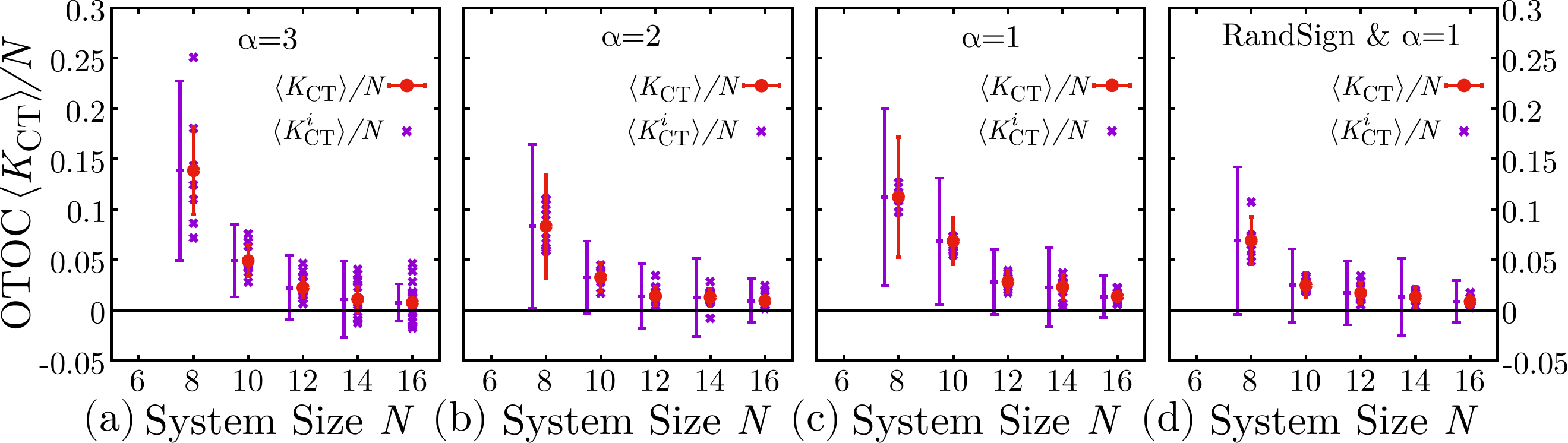}
\caption{Time average of the cross contribution to the number of correlated spins, $\langle K_{CT}\rangle/N$ (red solid circles), compared with time average of the cross term contribution of a single site, $\langle K_{CT}^{i}\rangle/N$ (purple crosses). Also shown the standard deviation from the average value $SD(K_{CT})/N$ (eq.\eqref{eqSDred}, red bars), and the square root of the site's variance average $\sqrt{\overline{\sigma_{CT}^{2}}}$ (eq. \eqref{eq:varsiteavg}, purple bars) which are slightly shifted for clarity purposes. (a) $D_{ij}\propto1/|r_{ij}|^{3}$ (b) $D_{ij}\propto1/|r_{ij}|^{2}$ and (c, d) $D_{ij}\propto1/|r_{ij}|$.}
\label{fig:03} 
\end{figure*}
\par\end{center}

In order to quantify the difference between the local and global OTOCs at long times, we computed the time average of the cross term at saturation $\langle K_{CT}^{i}\rangle=\frac{1}{\tau}\int_{t_{s}}^{t_{max}}K_{CT}^{i}(t){\rm {dt}}$, where $t_{max}$ is the final time in our simulation and $\tau=t_{max}-t_{s}$, for various systems sizes. In Fig.~\ref{fig:03}, we depict these magnitudes relative to the system size as a function of $N$, which can be interpreted as the relative error between $K_{G}$ and $K_{L}$, as $N$ is approximately their saturation value. The site contributions $\langle K_{CT}^{i}\rangle/N$ are shown with purple crosses, while their average $\langle K_{CT}\rangle/N$ is shown with red dots. In all cases we observe that not only, $\langle K_{CT}\rangle/N$ decreases with $N$ but also each $\langle K_{CT}^{i}\rangle/N$, implying that the equivalence of the global or local observation is valid for each individual initial state, without the necessity of summing all the initial conditions to have the effect. The error bars of the red dots in Fig. \ref{fig:03} correspond to the normalized standard deviation $SD(K_{CT})/N$, where 
\begin{equation}
SD(K_{CT})=\sqrt{\frac{1}{\tau}\int_{t_{s}}^{t_{max}}(K_{CT}(t)-\langle K_{CT}\rangle)^{2}{\rm {dt}}},\label{eqSDred}
\end{equation}
consequently quantifying the temporal fluctuations around the saturation value. An equivalent expression is used to define the standard deviation from the average value of each site $SD(K_{CT}^{i})/N$.

We observe that, just as $\langle K_{CT}^{i}\rangle/N$ decreases when $N$ increases, so do their fluctuations, as can be perceived in Fig.~\ref{fig:02}. The fluctuations in the long-time behavior of OTOCs in closed systems have been directly associated with chaos, i.e. the more chaotic the system the smaller the fluctuations in the long-time behavior of the OTOC~\cite{FGmJaWi19}. Given that the fluctuations of $K_{L}$ are considerably smaller than the fluctuations of $K_{G}$, we can state that $SD(K_{G})\approx SD(K_{CT})$. Consequently, as we increase $N$, the systems become more chaotic and the local and global OTOCs become almost identical. This is evidenced by the simultaneous decrease in the average value of $K_{CT}(t)$ and its fluctuations. This explains why in an NMR experiment, where there is a macroscopic number of spins involved, an environment and intrinsic experimental errors, this fluctuations can not be observed.

Moreover, we can extend the analysis to the magnitude of the fluctuations in individual and total OTOCs. We observe that fluctuations on $\langle K_{CT}\rangle$ are considerably smaller than the fluctuation of $\langle K_{CT}^{i}\rangle$. It is easy to see the relation between both magnitudes, 
\begin{eqnarray}
SD^{2}(K_{CT}) & = & \frac{1}{N^{2}}\sum_{i}SD^{2}(K_{CT}^{i})\nonumber \\
 & + & \frac{1}{N^{2}}\sum_{i\neq j}\mathrm{Cov}(K_{CT}^{i}(t),K_{CT}^{j}(t))\\
 & = & \frac{\overline{\sigma_{CT}^{2}}}{N}+\frac{1}{N^{2}}\sum_{i\neq j}\mathrm{Cov}(K_{CT}^{i}(t),K_{CT}^{j}(t))\nonumber 
\end{eqnarray}
where we denoted 
\begin{equation}
\overline{\sigma_{CT}^{2}}=\frac{1}{N}\sum_{i}SD^{2}(K_{CT}^{i}).\label{eq:varsiteavg}
\end{equation}
The over-line represents an average over initial sites (purple bars in Fig. \ref{fig:03}). Thus, if there were no correlation between different $K_{CT}^{i}(t)$, we would have $SD^{2}(K_{CT})=\overline{\sigma_{CT}^{2}}/N$. This last expression shows to be valid in the limit of large $N$, as can be seen in Appendix \ref{SM:Cov}. In fact, we see that the total correlation (sum of the covariances) decreases very rapidly with $N$. For $N=14$ they already become of the same order as our statistical precision. Furthermore, for the Hamiltonian including random signs in $D_{ij}$ the correlation between different $K_{CT}^{i}(t)$ this happens for $N=12$.

Notice that the dispersion between the values of $\langle K_{CT}^{i}\rangle/N$ is smaller when $\alpha$ is smaller. This can be rationalized by thinking that for large $\alpha$ the spins are less interconnected and site fluctuations, observed in each $\langle K_{CT}^{i}\rangle/N$, are highly dependent on the local fields $h_{i}$ affecting the spin and its neighbors.

\begin{figure}[t]
\centering \includegraphics[width=1\columnwidth]{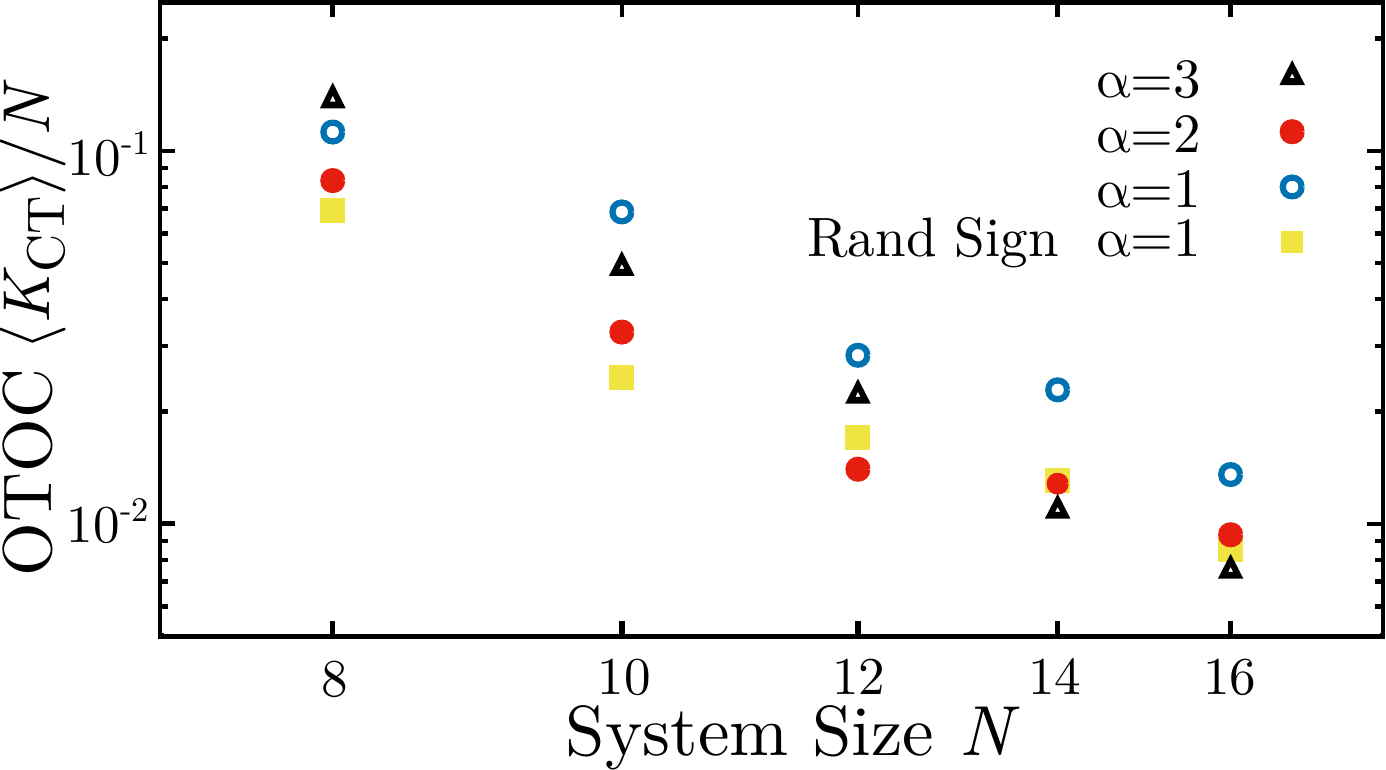}
\caption{Saturation value $\langle K_{CT}\rangle/N$ in log-log scale for spin rings interacting with Hamiltonian \eqref{eq:Ham}, with $D_{ij}=1/r_{ij}^{\alpha}$ for $\alpha={1,2,3}$. }
\label{fig:03b} 
\end{figure}

Fig.~\ref{fig:03b} shows the average value $\langle K_{CT}\rangle/N$ in logarithmic scale. It highlights the decay discussed above. The sizes are not large enough for the fittings to discriminate between an exponential decay or a power law. In the latter case, the exponent of this decay might vary between $-4.3$ and $-3.1$. In the case of complete random systems, an exponential decay is expected resulting from a homogeneous distribution of the states in the Hilbert space. In fact, for a DQ Hamiltonian we expect that regions of the Hilbert space corresponding to a total magnetization to have a normal distribution. Under this assumption, it is reasonable that the decay with $N$ follows a power law rather than an exponential. Nonetheless, the decay of the cross terms with $N$ leaves evidence that, in systems whose complexity is strong enough to generate chaotic dynamics, the global echoes are composed of a simple sum of local ones. Contributions from outside the original site will be completely uncorrelated (pseudo-random) at long times canceling each other. Thus, local and global OTOCs will provide the same information.

\section{Conclusions and Discussion.}

\label{Conclusions}

This work provides numerical evidence supporting the hypothesis that OTOCs obtained from global operators in NMR experiments are equivalent to an average over local operators OTOCs. Furthermore, this indicates that in a macroscopic homogeneous system, the global OTOCs observed in the experiment become representative of the single local OTOCs, which are of theoretical significance. Specifically, based on a theoretical analysis and numerical resolution of model systems, it states that global echoes and global OTOCs coincide with the average of local echoes and local OTOCs. This equivalence should be particularly valid as the complexity increases and the system size becomes macroscopic. However, as this limit is out of the possibilities of our classical computer, the numerical verification is restricted to fairly small systems. Nevertheless, in the numerical computations, we maintain the key ingredients of an experiment. Thus, the protocol and observables to extract the global OTOCs are the same as in an experimental setup. The local OTOCs, inaccessible experimentally, follow naturally from an expansion of the main observable in the experimental procedure, the total magnetization $\hat{I}^z$, as a sum of local spin polarizations ($\hat{I}^z=\sum_i \hat{I}^z_i$). Additionally, such calculation could model liquid crystal molecules, which were also observed experimentally~\cite{Sa+Ch16,SaLeACh09}. In this last case, the scrambling at long times through the whole Hilbert space acquires particular physical relevance as $K(t)$, the number of quantum mechanically correlated spins determined from the multiple quantum coherence distribution, becomes equal to the system size, adding further support to the spin counting procedure.

Our results show that the evolution of $K(t)$ has a similar behavior for the global $K_{G}(t)$ and local $K_{L}(t)$ observables. Indeed, after a brief initial time where they differ, both $K_{G}(t)$ and $K_{L}(t)$ grow together, exhibiting the same behavior over time. This is fundamental for assigning a local meaning to the global measurements, without being tied to the precise number of correlated spins. Their discrepancy, measured by $K_{CT}$, quickly becomes smaller than 10$\%$. The fact that after an initial transient, both the $K_{CT}$ values and their fluctuations go to zero with increasing system sizes, endorses the equivalence hypothesis. In our interpretation, the fundamental feature behind this correspondence is that the time reversal after a rotation cannot fully undo the many-body dynamics. This fact produces that a substantial number of backward paths in the Liouville space do not lead to the individual initial magnetization, but remain as multi-spin superposition without a net polarization. Thus, the observed polarization after a LE corresponds to small portions of paths that have unscrambled the multi-spin correlation into its original state.

Our conclusions are consistent, but not identical, with the results shown in Ref. \cite{Sw23}, by Zhou and Swingle. In that work, the expressions for the OTOCs discriminate the contribution of ``diagonal'' and ``non-diagonal'' correlations. The local OTOC ($K_{L}(t)$) determined in our work contains all the contributions identified as ``diagonal'' OTOCs in Ref. \cite{Sw23} along with some of their ``non-diagonal'' sums. These ``non-diagonal'' terms reflect the correlation between two evolutions that start from the same spin, but are disturbed at different sites. Although these contributions are generally negligible, they constitute a necessary conceptual difference to formulate an experimentally relevant vision in terms of local Loschmidt echoes.
The cross-correlation function ($K_{CT}(t)$) contains the remaining ``off-diagonal'' terms (Eq.~\eqref{eq:KCOTOC}). 
Thus, the assumption and results observed in a spin chain by Zhou and Swingle that the non-diagonal terms cancel out, implies the decay we found for $K_{CT}(t)$. This means that after an initial transient, the local OTOCs univocally determine the experimentally observed global OTOCs. 

The equivalence of global and local OTOCs is important for already available experimental platforms that allow determining information scrambling based on global control and observables. Thus, very diverse experimental results based on these approaches can now be interpreted as a probe of  the scrambling of local information. In particular, the results of this article support the hypothesis, explicitly stated in Ref.~\cite{Sa+Pa20}, that in both Loschmidt echoes and MQCs experiments, the evaluated observables over the whole system accurately reflect the ensemble average over uncorrelated ``local'' observables. This connection between MQC signal and local observables is crucial for the thermodynamics interpretation of several experiments in terms of local processes. Of particular importance are: 1) the ballistic growth of the number of entangled spins in a crystal under a DQ Hamiltonian \cite{Sa+Pa20}; 2) the diffusive scrambling of spin operators under transverse dipolar Hamiltonian, showing the relevant role of the many-body (Ising) terms in preventing ballistic propagation  \cite{Do-Al2021,DoAl21,SaPaCh23,ZwAl23}; 3) localization-delocalization transition of the controlled dynamics of quantum information in presence of perturbations built by mixing DQ and dipolar Hamiltonians, where ballistic and many-body terms compete, as a function of the mixture parameter \cite{ASu10,AlSu11a,Alvarez2013,ASuK15,Do-Al2021,DoAl21}; 4) in a more traditional context, the different excitation spreading rates observed in liquid crystal molecules before the saturation over the whole available Hilbert space is reached \cite{SaLeACh09,Sa+Ch16}.

While our results focus on NMR experimental conditions, they are also applicable to other experimental platforms where global control and observables are more easily achievable than their local counterparts, such as in experiments with trapped ions \cite{G+Rey17,Le+Re19} and ultra-cold polar molecules \cite{ZhSw23}.

\acknowledgments{Supercomputing time for this work was provided by CCAD (Centro de Computación de Alto Desempeño de la Universidad Nacional de Córdoba). GAA acknowledges support from CNEA; CONICET; ANPCyT-FONCyT PICT-2017-3156, PICT-2017-3699, PICT-2018-4333, PICT-2021-GRF-TI-00134, PICT-2021-I-A-00070; PIP-CONICET (11220170100486CO); UNCUYO SIIP Tipo I 2019-C028, 2022-C002, 2022-C030; Instituto Balseiro; Collaboration programs between the MINCyT (Argentina) and, MAECI (Italy) and MOST (Israel). Authors acknowledge to CONSOLIDAR SECYT-UNC 2018-2022, PIP-CONICET (11220200101508CO) and PICT-2017-2467 for their support. }

\appendix

\section{Multiple quantum coherences}\label{appA}

The $m$ order of coherence corresponds to matrix elements that represent transitions between many-spin states, in a Zeeman basis, that have different magnetization $m$. The evolution density matrix can be expressed by a superposition of contributions from different orders as 
\begin{equation}
\hat{\rho}=\sum_{m}\hat{\rho}_{m}\label{ec2}
\end{equation}

where the $m$-quantum coherence component behaves under rotation
as, 
\begin{equation}
e^{i\phi\hat{I}^{z}}\hat{\rho}_{m}e^{-i\phi\hat{I}^{z}}=\hat{\rho}_{m}e^{im\phi}.\label{ec1}
\end{equation}
Formally, the $m$ coherence intensity can be defined as 
\[
g_{m}=\frac{1}{\Tr\left\{(\hat{I}^{z})^{2}\right\}}\Tr\left\{\hat{\rho}_{m}\hat{\rho}_{-m}\right\}.
\]
Experimentally, by implementing systematic rotations around $z$ of angles $\phi$, the coherence distribution can be decoded through Fourier transformation of the collected signals, 
\[
M(\phi,t)=\frac{1}{\Tr\left\{(\hat{I}^{z})^{2}\right\}}\Tr\left\{e^{-i\phi \hat{I}^{z}}\hat{\rho}(t)e^{i\phi \hat{I}^{z}}e^{i\hat{\mathcal{H}}t}\hat{I}^{z}e^{-i\hat{\mathcal{H}}t}\right\},
\]
where $\phi$ is uniformly sample in steps of  $\Delta\phi=2\pi/2^M$,  with $ 2^M>m_{max}$ the maximum coherence order to be decoded. By expanding $\hat{\rho}(t)$ in the form \eqref{ec2}, considering $\hat{\rho}(0)=\hat{I}^{z}$, and using Eq. \eqref{ec1} (rotation property), the collected signals satisfy
\begin{eqnarray*}
    M(\phi,t)&=&\frac{1}{\Tr\left\{(\hat{I}^{z})^{2}\right\}}\Tr\left\{\sum_{m}\hat{\rho}_{m}e^{im\phi}\sum_{m}\hat{\rho}_{m}\right\}
    \\&=&\sum_{m}g_{m}e^{im\phi}.
\end{eqnarray*}
Note that, $M(\phi=0,t)=\sum_{m}g_{m}$ is the Loschmidt Echo intensity at $t$~\cite{SaPaCh23}. Separately, one can observe that the second moment of this MQC distribution is a global OTOC~\cite{Kh97}: 
\[
\begin{aligned}\sum_{m}m^{2}g_{m} & =-\left.\partial_{\phi}^{2}M(\phi,t)\right|_{\phi=0}\\
 & =\frac{1}{\Tr\left\{(\hat{I}^{z})^{2}\right\}}\Tr\left\{\left[\hat{I}^{z},\left[\hat{I}^{z},\hat{I}^{z}(t)\right]\right]\hat{I}^{z}(t)\right\}\\
 & =-\frac{1}{\Tr\left\{(\hat{I}^{z})^{2}\right\}}\Tr\left\{\left[\hat{I}^{z},\hat{I}^{z}(t)\right]\left[\hat{I}^{z},\hat{I}^{z}(t)\right]\right\}.
\end{aligned}
\]

\section{Experimental observables and qualitative comparison with numerical evaluations}\label{sm:expres}

\begin{figure}
\centering 
\includegraphics[width=1\columnwidth]{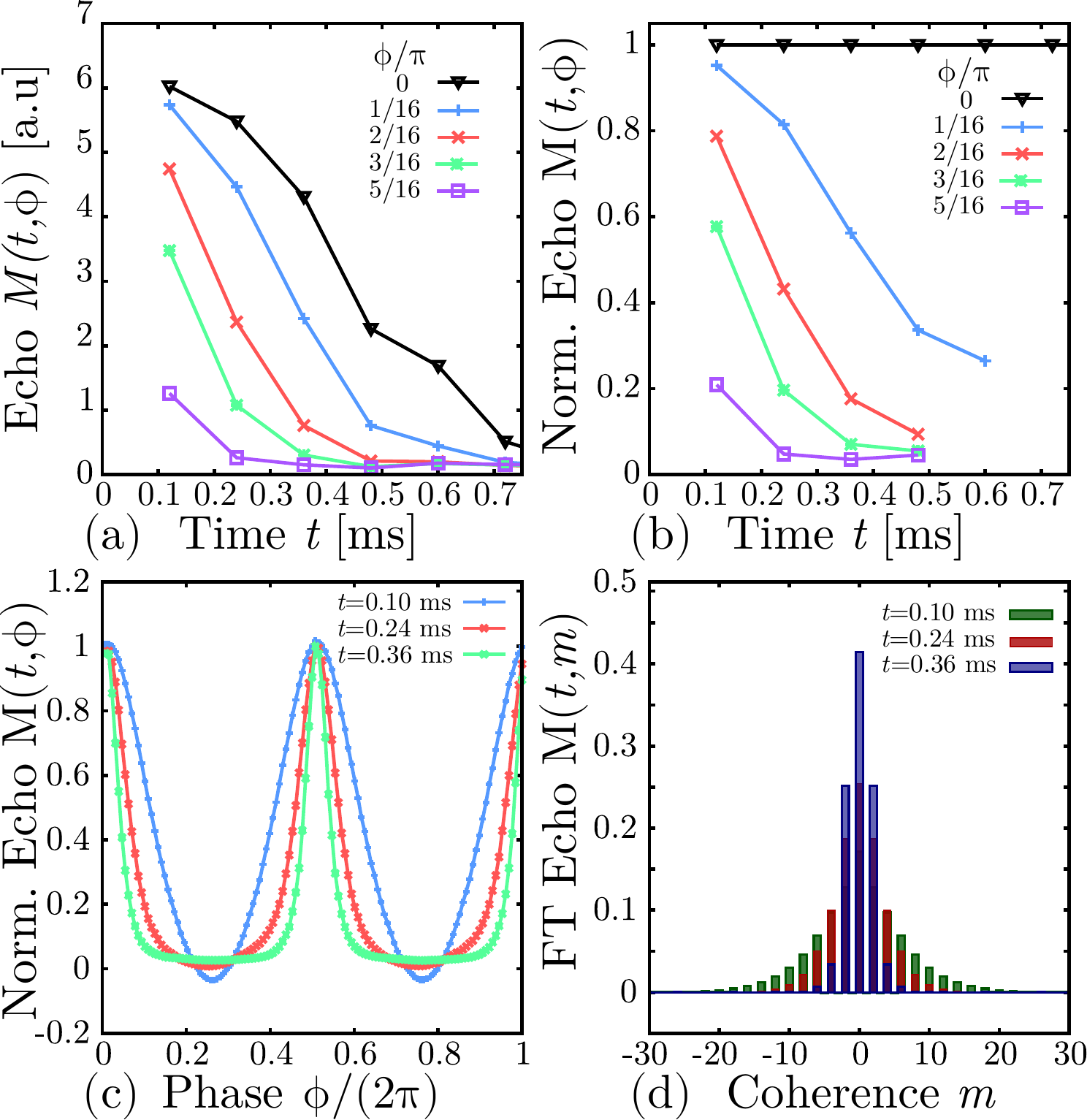}
\caption{(a) Experimental Loschmidt echo  $M(t,\phi)$. (b) Loschmidt Echo decay normalized by $M(t,0)$.
(c) Normalized LE as a function of $\phi$ at times $t=\{0.12, 0.24, 0.36\}$ ms. (d) Normalized coherence distribution (Fourier Transform of panel (c)). 
The data are part of the measurements carried out to prepare the results of the ref. \cite{SANCHEZ2023100104}.
}
\label{SM:fig:expresults} 
\end{figure}

Here, we provide a qualitative comparison between typical experimental observations and the simulated data presented in this article.  The experimental global echoes, determined from the total magnetization $I^z$ as given by Eq.~\eqref{echo}, are shown in Fig.~\ref{SM:fig:expresults}a, for various phases $\phi$, allowing comparison with the simulated curves in figure \ref{fig:NumImp}. 
The experimental data were obtained from 1H spins in an Adamantane crystal undergoing Double Quantum evolution. The pulse sequence used to achieve the average Hamiltonian consists of $16$ pulses, leading to a cycle time $t_c=0.12$ ms, with evolution times being multiples of $t_c$. These measurements are part of the data collected for the results of ref. \cite{SANCHEZ2023100104}.
In constrast to the numerical simulations, where $\phi=0$ ideally result in no decay, the experimental data exhibit a slow decay of the Loschmidt echo, attributed to small uncontrollable perturbations captured by $\hat{\Sigma}$. This decay typically follows a Gaussian trend at short times and an exponential behavior at longer times. To account for this, all echoes are renormalized, as shown in Fig.~\ref{SM:fig:expresults}b.
Fig.~\ref{SM:fig:expresults}c illustrates the normalized echo as a function of the phase $\phi$ for $t=\{0.12, 0.24, 0.36\}$ ms. Lastly, Fig.~\ref{SM:fig:expresults}d shows the distribution of multiple quantum coherence obtained from the Fourier transform of panel \((c)\).

\section{Mapping the local and global contributions to the cluster size K with diagonal and off-diagonal OTOCs}\label{appB}

In order to connect the local and global contributions to the number of correlated spins with OTOCs we start from definitions in Eqs. (\ref{MGLyC}, \ref{KGLyC}), namely, 
\begin{eqnarray}
M_{G}(t,\phi) & = & \frac{1}{N2^{N-2}}\Tr\left\{\hat{I}^{z}(t)\hat{R}^{\dagger}\hat{I}^{z}(t)\hat{R}\right\}\\
 & = & \frac{1}{N2^{N-2}}\sum_{i,j}\Tr\left\{\hat{I}_{i}^{z}(t)\hat{R}^{\dagger}\hat{I}_{j}^{z}(t)\hat{R}\right\}\nonumber \\
M_{L}(t,\phi) & = & \frac{1}{N2^{N-2}}\sum_{i}\Tr\left\{\hat{I}_{i}^{z}(t)\hat{R}^{\dagger}\hat{I}_{i}^{z}(t)\hat{R}\right\}\nonumber \\
M_{CT}(t,\phi) & = & \frac{1}{N2^{N-2}}\sum_{\substack{i,j\\
i\neq j
}
}\Tr\left\{\hat{I}_{i}^{z}(t)\hat{R}^{\dagger}\hat{I}_{j}^{z}(t)\hat{R}\right\}\nonumber .
\end{eqnarray}

We apply the second derivative to each term and analyze their contributions to OTOCs, 
\begin{eqnarray*}
\sum_{m}m^{2}g_{m}= & \hspace{-3cm}-\left.\partial_{\phi}^{2}M_{G}(\phi,t)\right|_{\phi=0}\\
= & -\left.\partial_{\phi}^{2}M_{L}(\phi,t)\right|_{\phi=0}-\left.\partial_{\phi}^{2}M_{CT}(\phi,t)\right|_{\phi=0},
\end{eqnarray*}
by doing so, we can explicitly write, from the echoes, $K_L$ and $K_{CT}$ as a combination of ``diagonal'' contributions of the form $\sum_{i,k}\Tr\left\{\left[\hat{I}_{k}^{z},\hat{I}_{i}^{z}(t)\right]^{2}\right\}$ and ``off-diagonal'' $\sum_{\substack{i,j,k,q\\
j\neq i\text{ or }k\neq q
}
}\Tr\left\{\left[\hat{I}_{q}^{z},\hat{I}_{i}^{z}(t)\right]\left[\hat{I}_{k}^{z},\hat{I}_{j}^{z}(t)\right]\right\}$ as defined in Ref. \cite{Sw23}. 

We found that, the only contribution of ``diagonal'' terms to the global OTOC comes from $M_{L}(t,\phi)$: 
\[
\begin{aligned}%
N2^{N-2}K_{L}(t) & =-2\left.\frac{\partial^{2}}{\partial\phi^{2}}M_{L}(t,\phi)\right|_{\phi=0}\\
 & =-2\sum_{i}\Tr\left\{\hat{I}_{i}^{z}(t)\hat{I}^{z}\hat{I}^{z}\hat{I}_{i}^{z}(t)-\hat{I}_{i}^{z}(t)\hat{I}^{z}\hat{I}_{i}^{z}(t)\hat{I}^{z}\right\}\\
 & =-2\sum_{i}\Tr\left\{\left[\hat{I}_{i}^{z},\left[\hat{I}^{z},\hat{I}_{i}^{z}(t)\right]\hat{I}_{i}^{z}(t)\right]\right\}\\
 & =-2\sum_{i}\Tr\left\{\left[\hat{I}^{z},\hat{I}_{i}^{z}(t)\right]^{2}\right\}\\
 & =-2\sum_{i,q,k}\Tr\left\{\left[\hat{I}_{q}^{z},\hat{I}_{i}^{z}(t)\right]\left[\hat{I}_{k}^{z},\hat{I}_{i}^{z}(t)\right]\right\}\\
 & =-2\left(\sum_{i,k}\Tr\left\{\left[\hat{I}_{k}^{z},\hat{I}_{i}^{z}(t)\right]^{2}\right\}\right.\\
 & \left.+\sum_{\substack{i,q,k\\
q\neq k
}
}\Tr\left\{\left[\hat{I}_{q}^{z},\hat{I}_{i}^{z}(t)\right]\left[\hat{I}_{k}^{z},\hat{I}_{i}^{z}(t)\right]\right\}\right),
\end{aligned}
\]
while only ``off-diagonal'' terms appear from cross-term $M_{CT}$:
\[
\begin{aligned}N2^{N-2}K_{CT}(t) & =-2\frac{\partial^{2}M_{CT}(t,\phi)}{\partial\phi^{2}}\\
 & =-2\sum_{i\neq j}\Tr\left\{\left[\hat{I}^{z},\hat{I}_{i}^{z}(t)\right]\left[\hat{I}^{z},\hat{I}_{j}^{z}(t)\right]\right\}\\
 & =-2\sum_{\substack{i,j,k,q\\
i\neq j
}
}\Tr\left\{\left[\hat{I}_{q}^{z},\hat{I}_{i}^{z}(t)\right]\left[\hat{I}_{k}^{z},\hat{I}_{j}^{z}(t)\right]\right\}.
\end{aligned}
\]

\begin{figure}[t]
\centering \includegraphics[width=1\columnwidth]{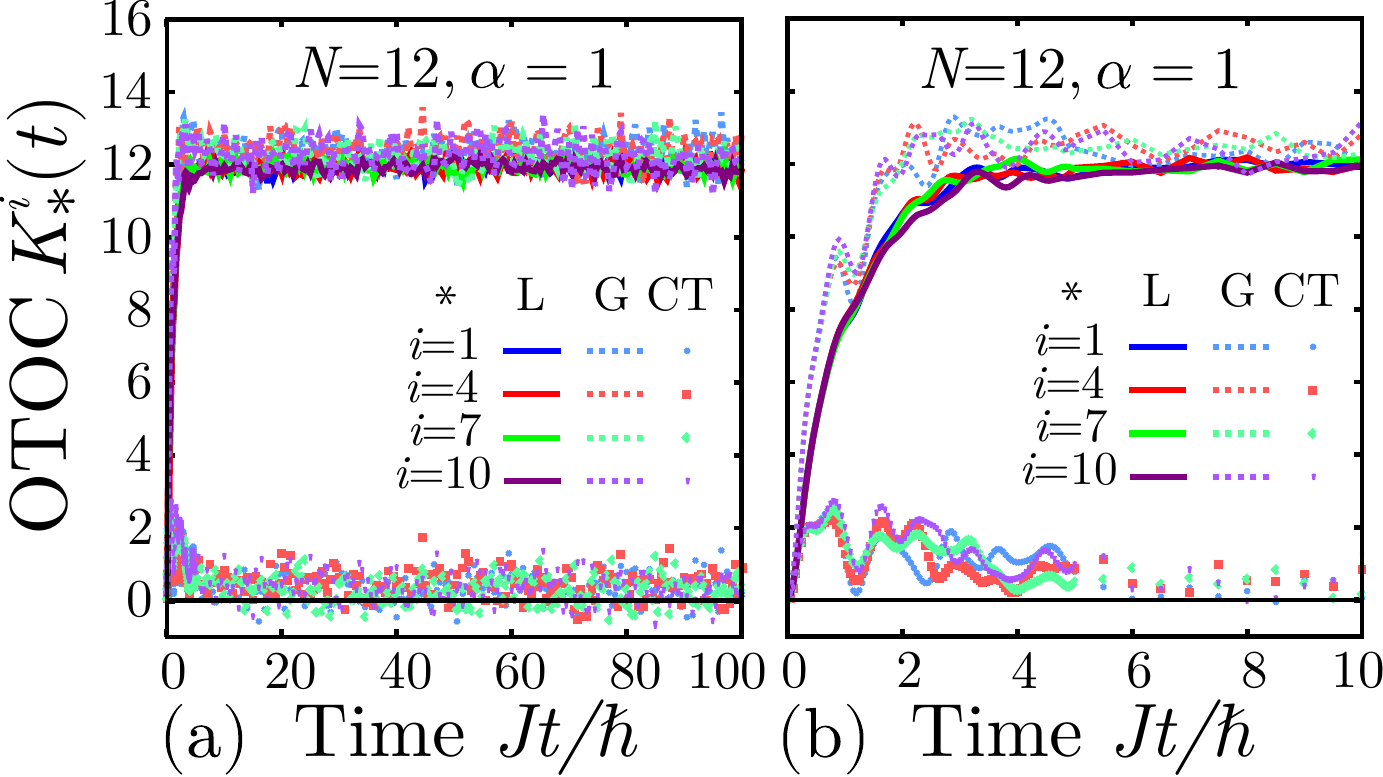}
\caption{Individual realizations per site, $K_{*}^{i}$, for a ring of spins with $N=12$ and $D_{ij}=J/|r_{i,j}|$. The supra-index $i$ represents the site, while the asterisk $*$ indicates whether the OTOC is Global, G, Local, L, or Cross Terms, CT. Panels (b) is a short-time zoom of the figures (a).}
\label{SM:fig:01} 
\end{figure}

\section{Short time behavior}\label{SM:ShortTime}

\begin{center}
\begin{figure*}
\centering \includegraphics[width=1\textwidth]{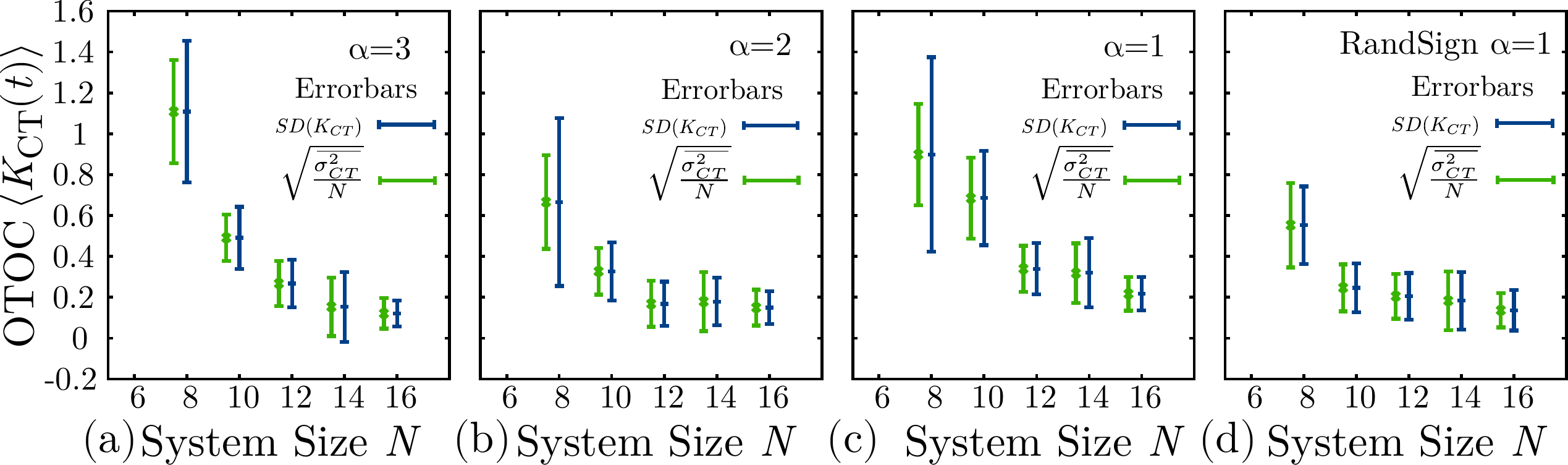} \caption{Time average of the cross contribution to the number of correlated spins, $\langle K_{CT}\rangle$. The standard deviation from the average value $SD(K_{CT})$ (Eq.~\eqref{eqSDred}, blue bars), and the square root of the site's variance average over $N$ ($\sqrt{\overline{\sigma_{CT}^{2}}/N}$, green bars) which are slightly shifted for clarity purposes. (a) $D_{ij}\propto1/|r_{ij}|^{3}$ (b) $D_{ij}\propto1/|r_{ij}|^{2}$ and (c) $D_{ij}\propto1/|r_{ij}|$. }
\label{SM:fig:cov} 
\end{figure*}
\par\end{center}

To derive the expression for the short-time behavior of $K_{G}(t)$, we start by using the Baker-Campbell-Hausdorff expansion in $\hat{I}^z(t)$, which approximates the time evolution of $\hat{I}^z$ under a Hamiltonian $\hat{\mathcal{H}}$: 
\begin{eqnarray}
\hat{I}^z(t) & \approx & \hat{I}^z+(-i\frac{t}{\hbar})[\hat{I}^z,\hat{\mathcal{H}}]\\
\left[\hat{I}^z,\hat{I}^z(t)\right] & \approx & \left[\hat{I}^z,\hat{I}^z+(-i\frac{t}{\hbar})[\hat{I}^z,\hat{\mathcal{H}}]\right]\\
 & \approx & (-i\frac{t}{\hbar})\left[\hat{I}^z,[\hat{I}^z,\hat{\mathcal{H}}]\right].
\end{eqnarray}

At this point, we carry out the commutator for double quantum Hamiltonian Eq. \eqref{eq:Ham}, which yields: 
\begin{eqnarray}
\left[\hat{I}^z,[\hat{I}^z,\hat{\mathcal{H}}_{DQ}]\right]=4\hbar^{2}\hat{\mathcal{H}}_{DQ}.
\end{eqnarray}

Finally, by substituting these expressions into Eq.~\eqref{eq:KGOTOC} and simplifying, we arrive at: 
\begin{eqnarray}
K_{G} & \approx & 2\frac{16t^{2}\hbar^{2}}{\Tr\left\{(\hat{I}^z)^{2}\right\}}\Tr\left\{\hat{\mathcal{H}}_{DQ}^{2}\right\}\\
 & = & \frac{16t^{2}\hbar^{2}}{\Tr\left\{(\hat{I}^z)^{2}\right\}}\sum_{i,j,k,l,i\neq j,k\neq l}D_{i,j}D_{k,l}\Tr\left\{\hat{\mathcal{H}}_{DQ_{i,j}}\hat{\mathcal{H}}_{DQ_{k,l}}\right\}\nonumber \\
 & = & 2\frac{16t^{2}\hbar^{2}}{\Tr\left\{(\hat{I}^z)^{2}\right\}}\sum_{i,j,k,l,i\neq j,k\neq l}2D_{i,j}D_{k,l}\Tr\left\{\hat{I}_{i}^{x}\hat{I}_{j}^{x}\hat{I}_{k}^{x}\hat{I}_{l}^{x}\right\}\nonumber \\
 & = & 2\frac{16t^{2}\hbar^{2}}{\Tr\left\{(\hat{I}^z)^{2}\right\}}\sum_{i\neq j}4D_{i,j}^{2}\Tr\left\{\hat{I}_{i}^{x}\hat{I}_{j}^{x}\hat{I}_{i}^{x}\hat{I}_{j}^{x}\right\}\nonumber \\
 & = & 2\frac{16t^{2}\hbar^{2}}{N2^{N-2}}\sum_{i\neq j}4D_{i,j}^{2}2^{N-4}\nonumber \\
 & = & \frac{32t^{2}\hbar^{2}}{N}\sum_{i,j,i\neq j}D_{i,j}^{2}.
\end{eqnarray}

Following the same procedure for a local OTOC we found that the initial growth only differs by a factor of two: 
\begin{eqnarray}
K_{L}(t) & = & -\frac{2}{N2^{N-2}}\sum_{i}\Tr\left\{\left[\hat{I}^{z},\hat{I}_{i}^{z}(t)\right]^{2}\right\}\\
 & \approx & \frac{2}{N2^{N-2}}16t^{2}\hbar^{2}\sum_{i,j}2D_{i,j}^{2}2^{N-4}\\
 & \approx & \frac{16}{N}t^{2}\hbar^{2}\sum_{i,j}D_{i,j}^{2}.
\end{eqnarray}

\section{Behavior of individual magnitudes \texorpdfstring{$K_{*}^{i}(t)$}{TEXT} and covariance.}\label{SM:Cov}

In the main text and preceding sections of the appendix, we have demonstrated that  $K_{*}(t)$ can be expressed as an average of site contributions, denoted as $K_{*}^{i}(t)$.
Each of these contributions exhibits minimal deviation from the averaged value $K_{*}(t)$, a fact supported by observing the variance of this average or directly comparing different curves, as depicted in Fig. \ref{SM:fig:01}. The curves corresponding to different initial sites differ mainly in fluctuations. Therefore, by averaging over initial sites, the primary effect is to mitigate these fluctuations, resulting in smoother curves.
Nevertheless, one can extract information about the spin correlation
\begin{eqnarray}
\langle K_{CT}\rangle & = & \frac{1}{N}\sum_{i}\langle K_{CT}^{i}\rangle\label{S17}\\
\langle K_{CT}^{2}\rangle & = & \frac{1}{N^{2}}\sum_{i,j}\langle K_{CT}^{i}K_{CT}^{j}\rangle\\
 & = & \frac{1}{N^{2}}\left[\sum_{i}\langle{K_{CT}^{i}}^{2}\rangle+\sum_{\substack{i,j\\
i\neq j
}
}\langle K_{CT}^{i}K_{CT}^{j}\rangle\right].
\end{eqnarray}

\begin{figure}
\centering \includegraphics[width=1\columnwidth]{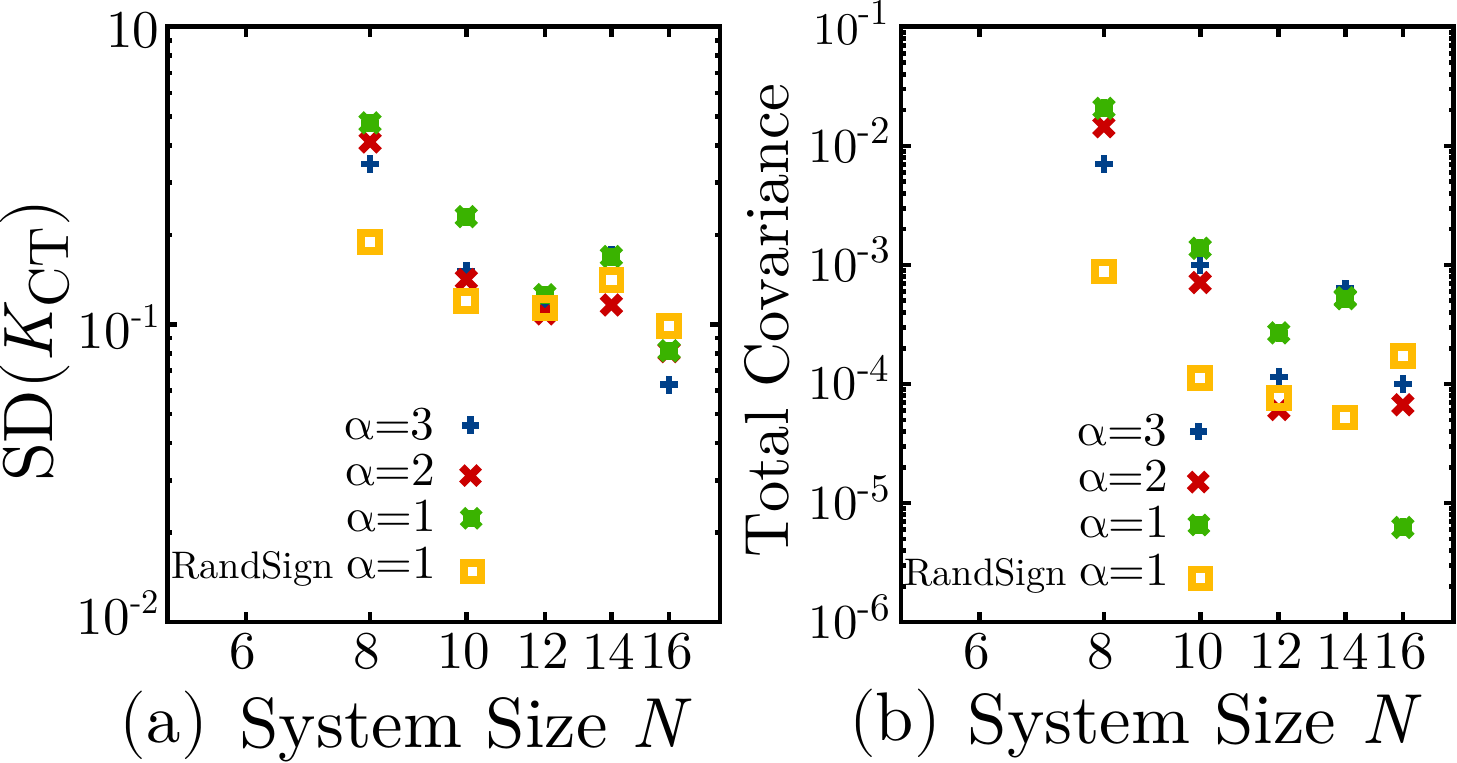}
\caption{(a) The standard deviation from the average value $SD(K_{CT})$ as
a function of $N$. (b) Total covariance (Eq. \eqref{SM:eq:TotCov})
as a function of $N$.}
\label{SM:fig:cov2} 
\end{figure}

By expanding Eq.~\eqref{eqSDred} into individual contributions, we
have, 
\begin{eqnarray*}
 & SD^{2}(K_{CT})=\\
 & \frac{1}{\tau N^{2}}\sum_{i,j}\int_{t_{s}}^{t_{max}}\left[K_{CT}^{i}(t)K_{CT}^{j}(t){\rm {dt}-\langle K_{CT}^{i}\rangle\langle K_{CT}^{j}\rangle}\right],
\end{eqnarray*}
which can be rearranged in the following form: 
\begin{eqnarray}
SD^{2}(K_{CT})=\frac{1}{N^{2}}\sum_{i}SD^{2}(K_{CT}^{i})\nonumber \\
+\frac{1}{\tau N^{2}}\sum_{\substack{i,j\\
i\neq j
}
}\int_{t_{s}}^{t_{max}}\left(K_{CT}^{i}(t)K_{CT}^{j}(t)-\langle K_{CT}^{i}\rangle\langle K_{CT}^{j}\rangle\right){\rm {dt}\nonumber}\\
=\frac{\overline{\sigma_{CT}^{2}}}{N}+\frac{1}{N^{2}}\sum_{i\neq j}\mathrm{Cov}(K_{CT}^{i},K_{CT}^{j}).\nonumber 
\end{eqnarray}
Here, we denoted 
\begin{equation}
\overline{\sigma_{CT}^{2}}=\frac{1}{N}\sum_{i}SD^{2}(K_{CT}^{i}),
\end{equation}
and define the total covariance as: 
\begin{equation}
\text{Total Cov.}=\frac{1}{N^{2}}\sum_{i\neq j}\mathrm{Cov}(K_{CT}^{i},K_{CT}^{j}).
\label{SM:eq:TotCov}
\end{equation}
this last term gives a measure of the total correlation between the dynamics of $K_{*}^{i}$. If the spin dynamics were uncorrelated, we would have $SD^{2}(K_{CT})=\frac{\overline{\sigma_{CT}^{2}}}{N}$. Fig. \ref{SM:fig:cov} compares these magnitudes for $K_{CT}$, we see that the error bars, representing $SD(K_{CT})$ and $\sqrt{\frac{\overline{\sigma_{CT}^{2}}}{N}}$ (blue and green bars respectively), becomes closer as $N$ increases. For a system with $\alpha=1$ plus random signs in the interactions this difference is small even for a small $N$.

Figure \ref{SM:fig:cov2}(a) shows the standard deviation $SD(K_{CT})$ as a function of $N$. As it was seen directly on the plots of $K_{*}(t)$ (Fig. \ref{fig:02}) the fluctuations decrease with the system size and are smaller when random signs are included in the Hamiltonian. Furthermore, this trend is also observed in the total covariance, Eq. \eqref{SM:eq:TotCov}, as it can be seen in Fig. \ref{SM:fig:cov2}(b). Indeed, this decrease is pronounced, becoming of the same order as our statistical precision for $N=14$.

\section{Loschmidt Echoes away from the initial site.}\label{sm:EchoAway}

\begin{figure}
\centering \includegraphics[width=1\columnwidth]{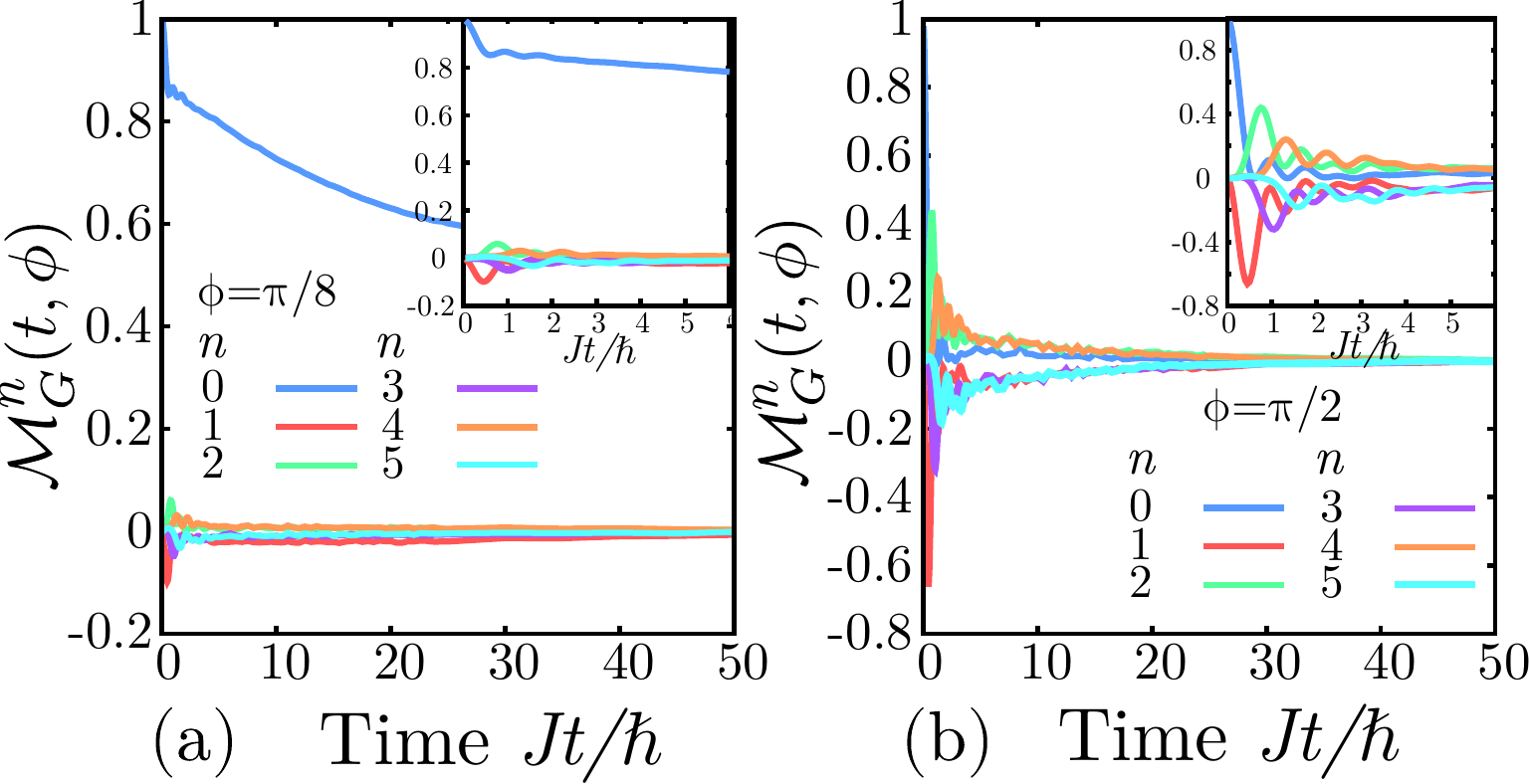}
\caption{Average return observed at a distance $n$ (colors) from the initial
site as a function of time. The inset show a zoom on short times.
Panel (a) correspond to perturbation value $\phi=\pi/8$ while panel
(b) to $\phi=\pi/2$. The numerical data correspond to a a ring of
$N=16$ with $\alpha=3$.}
\label{SM:fig:EchoByDistance} 
\end{figure}

As discussed in the main body of the paper, our numerical computations of the global and local Out-of-Time-Ordered Correlators (OTOCs) rely on the implementation of the MQC sequence. Fig. \ref{fig:NumImp} show typical results that, apart for the discrimination between local and global contributions, are similar to those usually found in experimental implementations.

In this section, we leverage the availability of numerical data and show the contributions to the total echo plots $M_{G}(t,\phi)$ from echoes observed at a distance $n$ from the initial site: $\mathcal{M}_{G}^{n}(t,\phi)=\frac{1}{N2^{N-2}}\sum_{i}\Tr\{(\hat{I}_{i+n}^{z}(t)+\hat{I}_{i-n}^{z}(t))\hat{R}^{\dagger}\hat{I}_{i}^{z}(t)\hat{R}\}(1-\delta_{n,0}/2)$.
Note that the contribution of sites $-n$ and $n$ comes from the ring geometry of the system, and it means a translation of $n$ sites to the right and left of site i. In the numerical implementation, the periodicity of the indexed needs to be done carefully. For example, for a ring of $N$ spins $\mathcal{M}_{G}^{0}(t,\phi)=\frac{1}{N2^{N-2}}\sum_{i}\Tr\{\hat{I}_{i}^{z}(t)\hat{R}^{\dagger}\hat{I}_{i}^{z}(t)\hat{R}\}$ and $\mathcal{M}_{G}^{1}(t,\phi)=\frac{1}{N2^{N-2}}\sum_{i}\Tr\{(\hat{I}_{i+1}^{z}(t)+\hat{I}_{i-1}^{z}(t))\hat{R}^{\dagger}\hat{I}_{i}^{z}(t)\hat{R}\}$, where $\hat{I}_{i+1}^{z}$ and $\hat{I}_{i-1}^{z}$ represent the spin at the right and left of $i$ respectively.

The behavior of these echoes are shown in Fig. \ref{SM:fig:EchoByDistance} for two different perturbations. One might think that, as occurs in the Polarization echo, what is missed from the original site may end up as polarization in the neighboring spins. However, in the DQ Hamiltonian it ends up as correlations which in general are not observable. Some perturbations, as $\phi=\pi/2$, could convert these correlations in observed magnetization at the neighboring sites. These correlations, however, are washed out much before the saturation times. It is clear how the system is highly correlated, observing a negative polarization returning at odd distances $n$, and positive polarization arriving at even values of $n$. This effect is clearer in Fig.~\ref{SM:fig:EchoByDistance}(b) due to the magnitude of the echoes. These echoes appear at longer times as $n$ increases. After this transient effect, we observed that all returning to $n\neq0$ go to zero, as already hinted in the previous analysis.

\bibliographystyle{apsrev4-2}
\bibliography{MQC-OTOC-NMR}

\begin{thebibliography}{80}%
\makeatletter
\providecommand \@ifxundefined [1]{%
 \@ifx{#1\undefined}
}%
\providecommand \@ifnum [1]{%
 \ifnum #1\expandafter \@firstoftwo
 \else \expandafter \@secondoftwo
 \fi
}%
\providecommand \@ifx [1]{%
 \ifx #1\expandafter \@firstoftwo
 \else \expandafter \@secondoftwo
 \fi
}%
\providecommand \natexlab [1]{#1}%
\providecommand \enquote  [1]{``#1''}%
\providecommand \bibnamefont  [1]{#1}%
\providecommand \bibfnamefont [1]{#1}%
\providecommand \citenamefont [1]{#1}%
\providecommand \href@noop [0]{\@secondoftwo}%
\providecommand \href [0]{\begingroup \@sanitize@url \@href}%
\providecommand \@href[1]{\@@startlink{#1}\@@href}%
\providecommand \@@href[1]{\endgroup#1\@@endlink}%
\providecommand \@sanitize@url [0]{\catcode `\\12\catcode `\$12\catcode
  `\&12\catcode `\#12\catcode `\^12\catcode `\_12\catcode `\%12\relax}%
\providecommand \@@startlink[1]{}%
\providecommand \@@endlink[0]{}%
\providecommand \url  [0]{\begingroup\@sanitize@url \@url }%
\providecommand \@url [1]{\endgroup\@href {#1}{\urlprefix }}%
\providecommand \urlprefix  [0]{URL }%
\providecommand \Eprint [0]{\href }%
\providecommand \doibase [0]{https://doi.org/}%
\providecommand \selectlanguage [0]{\@gobble}%
\providecommand \bibinfo  [0]{\@secondoftwo}%
\providecommand \bibfield  [0]{\@secondoftwo}%
\providecommand \translation [1]{[#1]}%
\providecommand \BibitemOpen [0]{}%
\providecommand \bibitemStop [0]{}%
\providecommand \bibitemNoStop [0]{.\EOS\space}%
\providecommand \EOS [0]{\spacefactor3000\relax}%
\providecommand \BibitemShut  [1]{\csname bibitem#1\endcsname}%
\let\auto@bib@innerbib\@empty
\bibitem [{\citenamefont {Goussev}\ \emph {et~al.}(2012)\citenamefont
  {Goussev}, \citenamefont {Jalabert}, \citenamefont {Pastawski},\ and\
  \citenamefont {Wisniacki}}]{GJaPaWi12}%
  \BibitemOpen
  \bibfield  {author} {\bibinfo {author} {\bibfnamefont {A.}~\bibnamefont
  {Goussev}}, \bibinfo {author} {\bibfnamefont {R.~A.}\ \bibnamefont
  {Jalabert}}, \bibinfo {author} {\bibfnamefont {H.~M.}\ \bibnamefont
  {Pastawski}},\ and\ \bibinfo {author} {\bibfnamefont {D.~A.}\ \bibnamefont
  {Wisniacki}},\ }\href {https://doi.org/10.4249/scholarpedia.11687} {\bibfield
   {journal} {\bibinfo  {journal} {Scholarpedia}\ }\textbf {\bibinfo {volume}
  {7}},\ \bibinfo {pages} {11687} (\bibinfo {year} {2012})}\BibitemShut
  {NoStop}%
\bibitem [{\citenamefont {Swingle}(2018)}]{Sw18}%
  \BibitemOpen
  \bibfield  {author} {\bibinfo {author} {\bibfnamefont {B.}~\bibnamefont
  {Swingle}},\ }\href {https://doi.org/10.1038/s41567-018-0295-5} {\bibfield
  {journal} {\bibinfo  {journal} {Nat. Phys.}\ }\textbf {\bibinfo {volume}
  {14}},\ \bibinfo {pages} {988} (\bibinfo {year} {2018})}\BibitemShut
  {NoStop}%
\bibitem [{\citenamefont {Lewis-Swan}\ \emph
  {et~al.}(2019{\natexlab{a}})\citenamefont {Lewis-Swan}, \citenamefont
  {Safavi-Naini}, \citenamefont {Kaufman},\ and\ \citenamefont
  {Rey}}]{LS+Rey19}%
  \BibitemOpen
  \bibfield  {author} {\bibinfo {author} {\bibfnamefont {R.~J.}\ \bibnamefont
  {Lewis-Swan}}, \bibinfo {author} {\bibfnamefont {A.}~\bibnamefont
  {Safavi-Naini}}, \bibinfo {author} {\bibfnamefont {A.~M.}\ \bibnamefont
  {Kaufman}},\ and\ \bibinfo {author} {\bibfnamefont {A.~M.}\ \bibnamefont
  {Rey}},\ }\href {https://doi.org/10.1038/s42254-019-0090-y} {\bibfield
  {journal} {\bibinfo  {journal} {Nat. Rev. Phys.}\ }\textbf {\bibinfo {volume}
  {1}},\ \bibinfo {pages} {627} (\bibinfo {year}
  {2019}{\natexlab{a}})}\BibitemShut {NoStop}%
\bibitem [{\citenamefont {Xu}\ and\ \citenamefont {Swingle}(2024)}]{Xu+Swi24}%
  \BibitemOpen
  \bibfield  {author} {\bibinfo {author} {\bibfnamefont {S.}~\bibnamefont
  {Xu}}\ and\ \bibinfo {author} {\bibfnamefont {B.}~\bibnamefont {Swingle}},\
  }\href {https://doi.org/10.1103/PRXQuantum.5.010201} {\bibfield  {journal}
  {\bibinfo  {journal} {PRX quantum}\ }\textbf {\bibinfo {volume} {5}},\
  \bibinfo {pages} {010201} (\bibinfo {year} {2024})}\BibitemShut {NoStop}%
\bibitem [{\citenamefont {Shenker}\ and\ \citenamefont
  {Stanford}(2014)}]{Shenker2014}%
  \BibitemOpen
  \bibfield  {author} {\bibinfo {author} {\bibfnamefont {S.~H.}\ \bibnamefont
  {Shenker}}\ and\ \bibinfo {author} {\bibfnamefont {D.}~\bibnamefont
  {Stanford}},\ }\href {https://doi.org/10.1007/JHEP03(2014)067} {\bibfield
  {journal} {\bibinfo  {journal} {Journal of High Energy Physics}\ }\textbf
  {\bibinfo {volume} {2014(3)}},\ \bibinfo {pages} {67} (\bibinfo {year}
  {2014})}\BibitemShut {NoStop}%
\bibitem [{\citenamefont {Bohrdt}\ \emph {et~al.}(2017)\citenamefont {Bohrdt},
  \citenamefont {Mendl}, \citenamefont {Endres},\ and\ \citenamefont
  {Knap}}]{BMEK17}%
  \BibitemOpen
  \bibfield  {author} {\bibinfo {author} {\bibfnamefont {A.}~\bibnamefont
  {Bohrdt}}, \bibinfo {author} {\bibfnamefont {C.~B.}\ \bibnamefont {Mendl}},
  \bibinfo {author} {\bibfnamefont {M.}~\bibnamefont {Endres}},\ and\ \bibinfo
  {author} {\bibfnamefont {M.}~\bibnamefont {Knap}},\ }\href
  {https://doi.org/10.1088/1367-2630/aa719b} {\bibfield  {journal} {\bibinfo
  {journal} {New J. Phys.}\ }\textbf {\bibinfo {volume} {19}},\ \bibinfo
  {pages} {063001} (\bibinfo {year} {2017})}\BibitemShut {NoStop}%
\bibitem [{\citenamefont {Maldacena}\ \emph {et~al.}(2016)\citenamefont
  {Maldacena}, \citenamefont {Shenker},\ and\ \citenamefont
  {Stanford}}]{MalSSt16}%
  \BibitemOpen
  \bibfield  {author} {\bibinfo {author} {\bibfnamefont {J.~M.}\ \bibnamefont
  {Maldacena}}, \bibinfo {author} {\bibfnamefont {S.~H.}\ \bibnamefont
  {Shenker}},\ and\ \bibinfo {author} {\bibfnamefont {D.}~\bibnamefont
  {Stanford}},\ }\href {https://doi.org/10.1007/JHEP08(2016)106} {\bibfield
  {journal} {\bibinfo  {journal} {Journal of High Energy Physics (Online)}\
  }\textbf {\bibinfo {volume} {2016(8)}},\ \bibinfo {pages} {106} (\bibinfo
  {year} {2016})}\BibitemShut {NoStop}%
\bibitem [{\citenamefont {Maldacena}(2020)}]{Mald20}%
  \BibitemOpen
  \bibfield  {author} {\bibinfo {author} {\bibfnamefont {J.~M.}\ \bibnamefont
  {Maldacena}},\ }\href {https://doi.org/10.1038/s42254-019-0146-z} {\bibfield
  {journal} {\bibinfo  {journal} {Nature Reviews Physics}\ }\textbf {\bibinfo
  {volume} {2}},\ \bibinfo {pages} {123} (\bibinfo {year} {2020})}\BibitemShut
  {NoStop}%
\bibitem [{\citenamefont {Kitaev}(2017)}]{Ki17}%
  \BibitemOpen
  \bibfield  {author} {\bibinfo {author} {\bibfnamefont {A.}~\bibnamefont
  {Kitaev}}} (\bibinfo {year} {2017}),\ \bibinfo {note} {in Brown Physics
  Colloquium, March 6}\BibitemShut {NoStop}%
\bibitem [{\citenamefont {{Larkin}}\ and\ \citenamefont
  {{Ovchinnikov}}(1969)}]{LaOv69}%
  \BibitemOpen
  \bibfield  {author} {\bibinfo {author} {\bibfnamefont {A.~I.}\ \bibnamefont
  {{Larkin}}}\ and\ \bibinfo {author} {\bibfnamefont {Y.~N.}\ \bibnamefont
  {{Ovchinnikov}}},\ }\href
  {http://www.jetp.ras.ru/cgi-bin/dn/e_028_06_1200.pdf} {\bibfield  {journal}
  {\bibinfo  {journal} {Soviet Journal of Experimental and Theoretical
  Physics}\ }\textbf {\bibinfo {volume} {28}},\ \bibinfo {pages} {1200}
  (\bibinfo {year} {1969})}\BibitemShut {NoStop}%
\bibitem [{\citenamefont {Hahn}(1950)}]{Hh50}%
  \BibitemOpen
  \bibfield  {author} {\bibinfo {author} {\bibfnamefont {E.}~\bibnamefont
  {Hahn}},\ }\href {https://doi.org/10.1103/PhysRev.80.580} {\bibfield
  {journal} {\bibinfo  {journal} {Phys. Rev.}\ }\textbf {\bibinfo {volume}
  {80}},\ \bibinfo {pages} {580} (\bibinfo {year} {1950})}\BibitemShut
  {NoStop}%
\bibitem [{\citenamefont {Rhim}\ \emph {et~al.}(1970)\citenamefont {Rhim},
  \citenamefont {Pines},\ and\ \citenamefont {Waugh}}]{RPiWa70}%
  \BibitemOpen
  \bibfield  {author} {\bibinfo {author} {\bibfnamefont {W.-K.}\ \bibnamefont
  {Rhim}}, \bibinfo {author} {\bibfnamefont {A.}~\bibnamefont {Pines}},\ and\
  \bibinfo {author} {\bibfnamefont {J.~S.}\ \bibnamefont {Waugh}},\ }\href
  {https://doi.org/10.1103/PhysRevLett.25.218} {\bibfield  {journal} {\bibinfo
  {journal} {Phys. Rev. Lett.}\ }\textbf {\bibinfo {volume} {25}},\ \bibinfo
  {pages} {218} (\bibinfo {year} {1970})}\BibitemShut {NoStop}%
\bibitem [{\citenamefont {Baum}\ \emph {et~al.}(1985)\citenamefont {Baum},
  \citenamefont {Munowitz}, \citenamefont {Garroway},\ and\ \citenamefont
  {Pines}}]{BMGPi85}%
  \BibitemOpen
  \bibfield  {author} {\bibinfo {author} {\bibfnamefont {J.}~\bibnamefont
  {Baum}}, \bibinfo {author} {\bibfnamefont {M.}~\bibnamefont {Munowitz}},
  \bibinfo {author} {\bibfnamefont {A.~N.}\ \bibnamefont {Garroway}},\ and\
  \bibinfo {author} {\bibfnamefont {A.}~\bibnamefont {Pines}},\ }\href
  {https://doi.org/10.1063/1.449344} {\bibfield  {journal} {\bibinfo  {journal}
  {J. Chem. Phys.}\ }\textbf {\bibinfo {volume} {83}},\ \bibinfo {pages} {2015}
  (\bibinfo {year} {1985})}\BibitemShut {NoStop}%
\bibitem [{\citenamefont {Munowitz}\ and\ \citenamefont
  {Pines}(1986)}]{MuPi86}%
  \BibitemOpen
  \bibfield  {author} {\bibinfo {author} {\bibfnamefont {M.}~\bibnamefont
  {Munowitz}}\ and\ \bibinfo {author} {\bibfnamefont {A.}~\bibnamefont
  {Pines}},\ }\href {https://doi.org/10.1126/science.233.4763.525} {\bibfield
  {journal} {\bibinfo  {journal} {Science}\ }\textbf {\bibinfo {volume}
  {233}},\ \bibinfo {pages} {525} (\bibinfo {year} {1986})}\BibitemShut
  {NoStop}%
\bibitem [{\citenamefont {Cho}\ \emph {et~al.}(2005)\citenamefont {Cho},
  \citenamefont {Ladd}, \citenamefont {Baugh}, \citenamefont {Cory},\ and\
  \citenamefont {Ramanathan}}]{C+CoRa05}%
  \BibitemOpen
  \bibfield  {author} {\bibinfo {author} {\bibfnamefont {H.}~\bibnamefont
  {Cho}}, \bibinfo {author} {\bibfnamefont {T.~D.}\ \bibnamefont {Ladd}},
  \bibinfo {author} {\bibfnamefont {J.}~\bibnamefont {Baugh}}, \bibinfo
  {author} {\bibfnamefont {D.~G.}\ \bibnamefont {Cory}},\ and\ \bibinfo
  {author} {\bibfnamefont {C.}~\bibnamefont {Ramanathan}},\ }\href
  {https://doi.org/10.1103/PhysRevB.72.054427} {\bibfield  {journal} {\bibinfo
  {journal} {Phys. Rev. B}\ }\textbf {\bibinfo {volume} {72}},\ \bibinfo
  {pages} {054427} (\bibinfo {year} {2005})}\BibitemShut {NoStop}%
\bibitem [{\citenamefont {Zhang}\ \emph {et~al.}(1992)\citenamefont {Zhang},
  \citenamefont {Meier},\ and\ \citenamefont {Ernst}}]{ZMEr92}%
  \BibitemOpen
  \bibfield  {author} {\bibinfo {author} {\bibfnamefont {S.}~\bibnamefont
  {Zhang}}, \bibinfo {author} {\bibfnamefont {B.~H.}\ \bibnamefont {Meier}},\
  and\ \bibinfo {author} {\bibfnamefont {R.~R.}\ \bibnamefont {Ernst}},\ }\href
  {https://doi.org/10.1103/physrevlett.69.2149} {\bibfield  {journal} {\bibinfo
   {journal} {Phys.Rev.Lett.}\ }\textbf {\bibinfo {volume} {69}},\ \bibinfo
  {pages} {2149} (\bibinfo {year} {1992})}\BibitemShut {NoStop}%
\bibitem [{\citenamefont {Levstein}\ \emph {et~al.}(1998)\citenamefont
  {Levstein}, \citenamefont {Usaj},\ and\ \citenamefont {Pastawski}}]{LUPa98}%
  \BibitemOpen
  \bibfield  {author} {\bibinfo {author} {\bibfnamefont {P.~R.}\ \bibnamefont
  {Levstein}}, \bibinfo {author} {\bibfnamefont {G.}~\bibnamefont {Usaj}},\
  and\ \bibinfo {author} {\bibfnamefont {H.~M.}\ \bibnamefont {Pastawski}},\
  }\href {https://doi.org/10.1063/1.475664} {\bibfield  {journal} {\bibinfo
  {journal} {J. Chem. Phys.}\ }\textbf {\bibinfo {volume} {108}},\ \bibinfo
  {pages} {2718} (\bibinfo {year} {1998})}\BibitemShut {NoStop}%
\bibitem [{\citenamefont {Usaj}\ \emph {et~al.}(1998)\citenamefont {Usaj},
  \citenamefont {Pastawski},\ and\ \citenamefont {Levstein}}]{UPaLe98}%
  \BibitemOpen
  \bibfield  {author} {\bibinfo {author} {\bibfnamefont {G.}~\bibnamefont
  {Usaj}}, \bibinfo {author} {\bibfnamefont {H.~M.}\ \bibnamefont
  {Pastawski}},\ and\ \bibinfo {author} {\bibfnamefont {P.~R.}\ \bibnamefont
  {Levstein}},\ }\href {https://doi.org/10.1080/00268979809483253} {\bibfield
  {journal} {\bibinfo  {journal} {Mol. Phys.}\ }\textbf {\bibinfo {volume}
  {95}},\ \bibinfo {pages} {1229} (\bibinfo {year} {1998})}\BibitemShut
  {NoStop}%
\bibitem [{\citenamefont {Jalabert}\ and\ \citenamefont
  {Pastawski}(2001)}]{JaPa01}%
  \BibitemOpen
  \bibfield  {author} {\bibinfo {author} {\bibfnamefont {R.~A.}\ \bibnamefont
  {Jalabert}}\ and\ \bibinfo {author} {\bibfnamefont {H.~M.}\ \bibnamefont
  {Pastawski}},\ }\href {https://doi.org/10.1103/PhysRevLett.86.2490}
  {\bibfield  {journal} {\bibinfo  {journal} {Phys. Rev. Lett.}\ }\textbf
  {\bibinfo {volume} {86}},\ \bibinfo {pages} {2490} (\bibinfo {year}
  {2001})}\BibitemShut {NoStop}%
\bibitem [{\citenamefont {{\'A}lvarez}\ and\ \citenamefont
  {Suter}(2010)}]{ASu10}%
  \BibitemOpen
  \bibfield  {author} {\bibinfo {author} {\bibfnamefont {G.~A.}\ \bibnamefont
  {{\'A}lvarez}}\ and\ \bibinfo {author} {\bibfnamefont {D.}~\bibnamefont
  {Suter}},\ }\href {https://doi.org/10.1103/PhysRevLett.104.230403} {\bibfield
   {journal} {\bibinfo  {journal} {Phys. Rev. Lett.}\ }\textbf {\bibinfo
  {volume} {104}},\ \bibinfo {pages} {230403} (\bibinfo {year}
  {2010})}\BibitemShut {NoStop}%
\bibitem [{\citenamefont {\'Alvarez}\ \emph {et~al.}(2015)\citenamefont
  {\'Alvarez}, \citenamefont {Suter},\ and\ \citenamefont {Kaiser}}]{ASuK15}%
  \BibitemOpen
  \bibfield  {author} {\bibinfo {author} {\bibfnamefont {G.~A.}\ \bibnamefont
  {\'Alvarez}}, \bibinfo {author} {\bibfnamefont {D.}~\bibnamefont {Suter}},\
  and\ \bibinfo {author} {\bibfnamefont {R.}~\bibnamefont {Kaiser}},\ }\href
  {https://doi.org/10.1126/science.1261160} {\bibfield  {journal} {\bibinfo
  {journal} {Science}\ }\textbf {\bibinfo {volume} {349}},\ \bibinfo {pages}
  {846} (\bibinfo {year} {2015})}\BibitemShut {NoStop}%
\bibitem [{\citenamefont {Gärttner}\ \emph {et~al.}(2017)\citenamefont
  {Gärttner}, \citenamefont {Bohnet}, \citenamefont {Wall}, \citenamefont
  {Bollinger},\ and\ \citenamefont {Rey}}]{G+Rey17}%
  \BibitemOpen
  \bibfield  {author} {\bibinfo {author} {\bibfnamefont {M.}~\bibnamefont
  {Gärttner}}, \bibinfo {author} {\bibfnamefont {A.}~\bibnamefont {Bohnet},
  \bibfnamefont {J.~G. Safavi-Naini}}, \bibinfo {author} {\bibfnamefont
  {M.~L.}\ \bibnamefont {Wall}}, \bibinfo {author} {\bibfnamefont {J.~J.}\
  \bibnamefont {Bollinger}},\ and\ \bibinfo {author} {\bibfnamefont {A.~M.}\
  \bibnamefont {Rey}},\ }\href {https://doi.org/10.1038/nphys4119} {\bibfield
  {journal} {\bibinfo  {journal} {Nature Physics}\ }\textbf {\bibinfo {volume}
  {13}},\ \bibinfo {pages} {781} (\bibinfo {year} {2017})}\BibitemShut
  {NoStop}%
\bibitem [{\citenamefont {Yao}\ and\ \citenamefont {Nayak}(2018)}]{YaNa18}%
  \BibitemOpen
  \bibfield  {author} {\bibinfo {author} {\bibfnamefont {N.~Y.}\ \bibnamefont
  {Yao}}\ and\ \bibinfo {author} {\bibfnamefont {C.}~\bibnamefont {Nayak}},\
  }\href {https://doi.org/10.1063/pt.3.4020} {\bibfield  {journal} {\bibinfo
  {journal} {Physics Today}\ }\textbf {\bibinfo {volume} {71}},\ \bibinfo
  {pages} {40} (\bibinfo {year} {2018})}\BibitemShut {NoStop}%
\bibitem [{\citenamefont {Wei}\ \emph {et~al.}(2018)\citenamefont {Wei},
  \citenamefont {Ramanathan},\ and\ \citenamefont {Cappellaro}}]{WeiChCa18}%
  \BibitemOpen
  \bibfield  {author} {\bibinfo {author} {\bibfnamefont {K.~X.}\ \bibnamefont
  {Wei}}, \bibinfo {author} {\bibfnamefont {C.}~\bibnamefont {Ramanathan}},\
  and\ \bibinfo {author} {\bibfnamefont {P.}~\bibnamefont {Cappellaro}},\
  }\href {https://doi.org/10.1103/PhysRevLett.120.070501} {\bibfield  {journal}
  {\bibinfo  {journal} {Phys. Rev. Lett.}\ }\textbf {\bibinfo {volume} {120}},\
  \bibinfo {pages} {070501} (\bibinfo {year} {2018})}\BibitemShut {NoStop}%
\bibitem [{\citenamefont {Lewis-Swan}\ \emph
  {et~al.}(2019{\natexlab{b}})\citenamefont {Lewis-Swan}, \citenamefont
  {Safavi-Naini}, \citenamefont {Bollinger},\ and\ \citenamefont
  {Rey}}]{LS+BolRey19}%
  \BibitemOpen
  \bibfield  {author} {\bibinfo {author} {\bibfnamefont {R.~J.}\ \bibnamefont
  {Lewis-Swan}}, \bibinfo {author} {\bibfnamefont {A.}~\bibnamefont
  {Safavi-Naini}}, \bibinfo {author} {\bibfnamefont {J.~J.}\ \bibnamefont
  {Bollinger}},\ and\ \bibinfo {author} {\bibfnamefont {A.~M.}\ \bibnamefont
  {Rey}},\ }\href {https://doi.org/10.1038/s41467-019-09436-y} {\bibfield
  {journal} {\bibinfo  {journal} {Nat. Comm.}\ }\textbf {\bibinfo {volume}
  {10}},\ \bibinfo {pages} {1581} (\bibinfo {year}
  {2019}{\natexlab{b}})}\BibitemShut {NoStop}%
\bibitem [{\citenamefont {Wei}\ \emph {et~al.}(2019)\citenamefont {Wei},
  \citenamefont {Peng}, \citenamefont {Shtanko}, \citenamefont {Marvian},
  \citenamefont {Lloyd}, \citenamefont {Ramanathan},\ and\ \citenamefont
  {Cappellaro}}]{Wei+Ca19}%
  \BibitemOpen
  \bibfield  {author} {\bibinfo {author} {\bibfnamefont {K.~X.}\ \bibnamefont
  {Wei}}, \bibinfo {author} {\bibfnamefont {P.}~\bibnamefont {Peng}}, \bibinfo
  {author} {\bibfnamefont {O.}~\bibnamefont {Shtanko}}, \bibinfo {author}
  {\bibfnamefont {I.}~\bibnamefont {Marvian}}, \bibinfo {author} {\bibfnamefont
  {S.}~\bibnamefont {Lloyd}}, \bibinfo {author} {\bibfnamefont
  {C.}~\bibnamefont {Ramanathan}},\ and\ \bibinfo {author} {\bibfnamefont
  {P.}~\bibnamefont {Cappellaro}},\ }\href
  {https://doi.org/10.1103/PhysRevLett.123.090605} {\bibfield  {journal}
  {\bibinfo  {journal} {Phys. Rev. Lett.}\ }\textbf {\bibinfo {volume} {123}},\
  \bibinfo {pages} {090605} (\bibinfo {year} {2019})}\BibitemShut {NoStop}%
\bibitem [{\citenamefont {S\'anchez}\ \emph {et~al.}(2020)\citenamefont
  {S\'anchez}, \citenamefont {Chattah}, \citenamefont {Wei}, \citenamefont
  {Buljubasich}, \citenamefont {Cappellaro},\ and\ \citenamefont
  {Pastawski}}]{Sa+Pa20}%
  \BibitemOpen
  \bibfield  {author} {\bibinfo {author} {\bibfnamefont {C.~M.}\ \bibnamefont
  {S\'anchez}}, \bibinfo {author} {\bibfnamefont {A.~K.}\ \bibnamefont
  {Chattah}}, \bibinfo {author} {\bibfnamefont {K.~X.}\ \bibnamefont {Wei}},
  \bibinfo {author} {\bibfnamefont {L.}~\bibnamefont {Buljubasich}}, \bibinfo
  {author} {\bibfnamefont {P.}~\bibnamefont {Cappellaro}},\ and\ \bibinfo
  {author} {\bibfnamefont {H.~M.}\ \bibnamefont {Pastawski}},\ }\href
  {https://doi.org/10.1103/PhysRevLett.124.030601} {\bibfield  {journal}
  {\bibinfo  {journal} {Phys. Rev. Lett.}\ }\textbf {\bibinfo {volume} {124}},\
  \bibinfo {pages} {030601} (\bibinfo {year} {2020})}\BibitemShut {NoStop}%
\bibitem [{\citenamefont {Niknam}\ \emph {et~al.}(2020)\citenamefont {Niknam},
  \citenamefont {Santos},\ and\ \citenamefont {Cory}}]{NSCo20}%
  \BibitemOpen
  \bibfield  {author} {\bibinfo {author} {\bibfnamefont {M.}~\bibnamefont
  {Niknam}}, \bibinfo {author} {\bibfnamefont {L.~F.}\ \bibnamefont {Santos}},\
  and\ \bibinfo {author} {\bibfnamefont {D.~G.}\ \bibnamefont {Cory}},\ }\href
  {https://doi.org/10.1103/PhysRevResearch.2.013200} {\bibfield  {journal}
  {\bibinfo  {journal} {Phys. Rev. Research}\ }\textbf {\bibinfo {volume}
  {2}},\ \bibinfo {pages} {013200} (\bibinfo {year} {2020})}\BibitemShut
  {NoStop}%
\bibitem [{\citenamefont {Yan}\ \emph {et~al.}(2020)\citenamefont {Yan},
  \citenamefont {Cincio},\ and\ \citenamefont {Zurek}}]{YCZu20}%
  \BibitemOpen
  \bibfield  {author} {\bibinfo {author} {\bibfnamefont {B.}~\bibnamefont
  {Yan}}, \bibinfo {author} {\bibfnamefont {L.}~\bibnamefont {Cincio}},\ and\
  \bibinfo {author} {\bibfnamefont {W.~H.}\ \bibnamefont {Zurek}},\ }\href
  {https://doi.org/10.1103/physrevlett.124.160603} {\bibfield  {journal}
  {\bibinfo  {journal} {Phys. Rev. Lett.}\ }\textbf {\bibinfo {volume} {124}},\
  \bibinfo {pages} {160603} (\bibinfo {year} {2020})}\BibitemShut {NoStop}%
\bibitem [{\citenamefont {Dom\'{\i}nguez}\ \emph {et~al.}(2021)\citenamefont
  {Dom\'{\i}nguez}, \citenamefont {Rodr\'{\i}guez}, \citenamefont {Kaiser},
  \citenamefont {Suter},\ and\ \citenamefont {\'Alvarez}}]{Do-Al2021}%
  \BibitemOpen
  \bibfield  {author} {\bibinfo {author} {\bibfnamefont {F.~D.}\ \bibnamefont
  {Dom\'{\i}nguez}}, \bibinfo {author} {\bibfnamefont {M.~C.}\ \bibnamefont
  {Rodr\'{\i}guez}}, \bibinfo {author} {\bibfnamefont {R.}~\bibnamefont
  {Kaiser}}, \bibinfo {author} {\bibfnamefont {D.}~\bibnamefont {Suter}},\ and\
  \bibinfo {author} {\bibfnamefont {G.~A.}\ \bibnamefont {\'Alvarez}},\ }\href
  {https://doi.org/10.1103/PhysRevA.104.012402} {\bibfield  {journal} {\bibinfo
   {journal} {Phys. Rev. A}\ }\textbf {\bibinfo {volume} {104}},\ \bibinfo
  {pages} {012402} (\bibinfo {year} {2021})}\BibitemShut {NoStop}%
\bibitem [{\citenamefont {Zhou}\ and\ \citenamefont
  {Swingle}(2023{\natexlab{a}})}]{Sw23}%
  \BibitemOpen
  \bibfield  {author} {\bibinfo {author} {\bibfnamefont {T.}~\bibnamefont
  {Zhou}}\ and\ \bibinfo {author} {\bibfnamefont {B.}~\bibnamefont {Swingle}},\
  }\href {https://doi.org/10.1038/s41467-023-39065-5} {\bibfield  {journal}
  {\bibinfo  {journal} {Natture Communications}\ }\textbf {\bibinfo {volume}
  {14}},\ \bibinfo {pages} {3411} (\bibinfo {year}
  {2023}{\natexlab{a}})}\BibitemShut {NoStop}%
\bibitem [{\citenamefont {Álvarez}\ and\ \citenamefont
  {Suter}(2011)}]{AlSu11}%
  \BibitemOpen
  \bibfield  {author} {\bibinfo {author} {\bibfnamefont {G.~A.}\ \bibnamefont
  {Álvarez}}\ and\ \bibinfo {author} {\bibfnamefont {D.}~\bibnamefont
  {Suter}},\ }\href {https://doi.org/10.1103/PhysRevLett.107.230501} {\bibfield
   {journal} {\bibinfo  {journal} {Phys. Rev. Lett.}\ }\textbf {\bibinfo
  {volume} {107}},\ \bibinfo {pages} {230501} (\bibinfo {year}
  {2011})}\BibitemShut {NoStop}%
\bibitem [{\citenamefont {Maurer}\ \emph {et~al.}(2012)\citenamefont {Maurer},
  \citenamefont {Kucsko}, \citenamefont {Latta}, \citenamefont {Jiang},
  \citenamefont {Yao}, \citenamefont {Bennett}, \citenamefont {Pastawski},
  \citenamefont {Hunger}, \citenamefont {Chisholm}, \citenamefont {Markham},
  \citenamefont {Twitchen}, \citenamefont {Cirac},\ and\ \citenamefont
  {Lukin}}]{Ma+PaLu12}%
  \BibitemOpen
  \bibfield  {author} {\bibinfo {author} {\bibfnamefont {P.~C.}\ \bibnamefont
  {Maurer}}, \bibinfo {author} {\bibfnamefont {G.}~\bibnamefont {Kucsko}},
  \bibinfo {author} {\bibfnamefont {C.}~\bibnamefont {Latta}}, \bibinfo
  {author} {\bibfnamefont {L.}~\bibnamefont {Jiang}}, \bibinfo {author}
  {\bibfnamefont {N.~Y.}\ \bibnamefont {Yao}}, \bibinfo {author} {\bibfnamefont
  {S.~D.}\ \bibnamefont {Bennett}}, \bibinfo {author} {\bibfnamefont
  {F.}~\bibnamefont {Pastawski}}, \bibinfo {author} {\bibfnamefont
  {D.}~\bibnamefont {Hunger}}, \bibinfo {author} {\bibfnamefont
  {N.}~\bibnamefont {Chisholm}}, \bibinfo {author} {\bibfnamefont
  {M.}~\bibnamefont {Markham}}, \bibinfo {author} {\bibfnamefont {D.~J.}\
  \bibnamefont {Twitchen}}, \bibinfo {author} {\bibfnamefont {J.~I.}\
  \bibnamefont {Cirac}},\ and\ \bibinfo {author} {\bibfnamefont {M.~D.}\
  \bibnamefont {Lukin}},\ }\href {https://doi.org/10.1126/science.1220513}
  {\bibfield  {journal} {\bibinfo  {journal} {Science}\ }\textbf {\bibinfo
  {volume} {336}},\ \bibinfo {pages} {1283} (\bibinfo {year}
  {2012})}\BibitemShut {NoStop}%
\bibitem [{\citenamefont {Souza}\ \emph {et~al.}(2012)\citenamefont {Souza},
  \citenamefont {{\'A}lvarez},\ and\ \citenamefont {Suter}}]{SoAlSu12}%
  \BibitemOpen
  \bibfield  {author} {\bibinfo {author} {\bibfnamefont {A.~M.}\ \bibnamefont
  {Souza}}, \bibinfo {author} {\bibfnamefont {G.~A.}\ \bibnamefont
  {{\'A}lvarez}},\ and\ \bibinfo {author} {\bibfnamefont {D.}~\bibnamefont
  {Suter}},\ }\href {https://doi.org/10.1098/rsta.2011.0355} {\bibfield
  {journal} {\bibinfo  {journal} {Philosophical Transactions of the Royal
  Society A: Mathematical, Physical and Engineering Sciences}\ }\textbf
  {\bibinfo {volume} {370}},\ \bibinfo {pages} {4748} (\bibinfo {year}
  {2012})}\BibitemShut {NoStop}%
\bibitem [{\citenamefont {Suter}\ and\ \citenamefont
  {{\'A}lvarez}(2016)}]{SuAl16}%
  \BibitemOpen
  \bibfield  {author} {\bibinfo {author} {\bibfnamefont {D.}~\bibnamefont
  {Suter}}\ and\ \bibinfo {author} {\bibfnamefont {G.~A.}\ \bibnamefont
  {{\'A}lvarez}},\ }\href {https://doi.org/10.1103/RevModPhys.88.041001}
  {\bibfield  {journal} {\bibinfo  {journal} {Rev. Mod. Phys.}\ ,\ \bibinfo
  {pages} {041001}} (\bibinfo {year} {2016})}\BibitemShut {NoStop}%
\bibitem [{\citenamefont {S\'anchez}\ \emph {et~al.}(2016)\citenamefont
  {S\'anchez}, \citenamefont {Levstein}, \citenamefont {Buljubasich},
  \citenamefont {Pastawski},\ and\ \citenamefont {Chattah}}]{Sa+Ch16}%
  \BibitemOpen
  \bibfield  {author} {\bibinfo {author} {\bibfnamefont {C.~M.}\ \bibnamefont
  {S\'anchez}}, \bibinfo {author} {\bibfnamefont {P.~R.}\ \bibnamefont
  {Levstein}}, \bibinfo {author} {\bibfnamefont {L.}~\bibnamefont
  {Buljubasich}}, \bibinfo {author} {\bibfnamefont {H.~M.}\ \bibnamefont
  {Pastawski}},\ and\ \bibinfo {author} {\bibfnamefont {A.~K.}\ \bibnamefont
  {Chattah}},\ }\href {https://doi.org/10.1098/rsta.2015.0155} {\bibfield
  {journal} {\bibinfo  {journal} {Phil. Trans. R. Soc. A}\ }\textbf {\bibinfo
  {volume} {374}},\ \bibinfo {pages} {20150155} (\bibinfo {year}
  {2016})}\BibitemShut {NoStop}%
\bibitem [{\citenamefont {Dom\'{\i}nguez}\ and\ \citenamefont
  {\'Alvarez}(2021)}]{DoAl21}%
  \BibitemOpen
  \bibfield  {author} {\bibinfo {author} {\bibfnamefont {F.~D.}\ \bibnamefont
  {Dom\'{\i}nguez}}\ and\ \bibinfo {author} {\bibfnamefont {G.~A.}\
  \bibnamefont {\'Alvarez}},\ }\href
  {https://doi.org/10.1103/PhysRevA.104.062406} {\bibfield  {journal} {\bibinfo
   {journal} {Phys. Rev. A}\ }\textbf {\bibinfo {volume} {104}},\ \bibinfo
  {pages} {062406} (\bibinfo {year} {2021})}\BibitemShut {NoStop}%
\bibitem [{\citenamefont {S\'anchez}\ \emph {et~al.}(2022)\citenamefont
  {S\'anchez}, \citenamefont {Chattah},\ and\ \citenamefont
  {Pastawski}}]{SaChPa22}%
  \BibitemOpen
  \bibfield  {author} {\bibinfo {author} {\bibfnamefont {C.~M.}\ \bibnamefont
  {S\'anchez}}, \bibinfo {author} {\bibfnamefont {A.~K.}\ \bibnamefont
  {Chattah}},\ and\ \bibinfo {author} {\bibfnamefont {H.~M.}\ \bibnamefont
  {Pastawski}},\ }\href {https://doi.org/10.1103/PhysRevA.105.052232}
  {\bibfield  {journal} {\bibinfo  {journal} {Phys. Rev. A}\ }\textbf {\bibinfo
  {volume} {105}},\ \bibinfo {pages} {052232} (\bibinfo {year}
  {2022})}\BibitemShut {NoStop}%
\bibitem [{\citenamefont {Sánchez}\ \emph
  {et~al.}(2023{\natexlab{a}})\citenamefont {Sánchez}, \citenamefont
  {Pastawski},\ and\ \citenamefont {Chattah}}]{SaPaCh23}%
  \BibitemOpen
  \bibfield  {author} {\bibinfo {author} {\bibfnamefont {C.}~\bibnamefont
  {Sánchez}}, \bibinfo {author} {\bibfnamefont {H.}~\bibnamefont
  {Pastawski}},\ and\ \bibinfo {author} {\bibfnamefont {A.}~\bibnamefont
  {Chattah}},\ }\href
  {https://doi.org/https://doi.org/10.1016/j.jmro.2023.100104} {\bibfield
  {journal} {\bibinfo  {journal} {Journal of Magnetic Resonance Open}\ }\textbf
  {\bibinfo {volume} {16-17}},\ \bibinfo {pages} {100104} (\bibinfo {year}
  {2023}{\natexlab{a}})}\BibitemShut {NoStop}%
\bibitem [{\citenamefont {Zwick}\ and\ \citenamefont
  {Álvarez}(2023)}]{ZwAl23}%
  \BibitemOpen
  \bibfield  {author} {\bibinfo {author} {\bibfnamefont {A.}~\bibnamefont
  {Zwick}}\ and\ \bibinfo {author} {\bibfnamefont {G.~A.}\ \bibnamefont
  {Álvarez}},\ }\href {https://doi.org/10.1016/j.jmro.2023.100113} {\bibfield
  {journal} {\bibinfo  {journal} {Journal of Magnetic Resonance Open}\ }\textbf
  {\bibinfo {volume} {16-17}},\ \bibinfo {pages} {100113} (\bibinfo {year}
  {2023})}\BibitemShut {NoStop}%
\bibitem [{\citenamefont {Levstein}\ \emph {et~al.}(2004)\citenamefont
  {Levstein}, \citenamefont {Chattah}, \citenamefont {Pastawski}, \citenamefont
  {Raya},\ and\ \citenamefont {Hirschinger}}]{LeChPa04}%
  \BibitemOpen
  \bibfield  {author} {\bibinfo {author} {\bibfnamefont {P.~R.}\ \bibnamefont
  {Levstein}}, \bibinfo {author} {\bibfnamefont {A.~K.}\ \bibnamefont
  {Chattah}}, \bibinfo {author} {\bibfnamefont {H.~M.}\ \bibnamefont
  {Pastawski}}, \bibinfo {author} {\bibfnamefont {J.}~\bibnamefont {Raya}},\
  and\ \bibinfo {author} {\bibfnamefont {J.}~\bibnamefont {Hirschinger}},\
  }\href {https://doi.org/10.1063/1.1792575} {\bibfield  {journal} {\bibinfo
  {journal} {J. Chem. Phys.}\ }\textbf {\bibinfo {volume} {121}},\ \bibinfo
  {pages} {7313} (\bibinfo {year} {2004})}\BibitemShut {NoStop}%
\bibitem [{\citenamefont {Kuffer}\ \emph {et~al.}(2024)\citenamefont {Kuffer},
  \citenamefont {Zwick},\ and\ \citenamefont {Álvarez}}]{KuZwAl24}%
  \BibitemOpen
  \bibfield  {author} {\bibinfo {author} {\bibfnamefont {M.}~\bibnamefont
  {Kuffer}}, \bibinfo {author} {\bibfnamefont {A.}~\bibnamefont {Zwick}},\ and\
  \bibinfo {author} {\bibfnamefont {G.~A.}\ \bibnamefont {Álvarez}},\ }\href
  {http://arxiv.org/abs/2405.04742} {\bibinfo {title} {Sensing
  {Out}-of-{Equilibrium} and {Quantum} {Non}-{Gaussian} environments via
  induced {Time}-{Reversal} {Symmetry} {Breaking} on the quantum-probe
  dynamics}} (\bibinfo {year} {2024}),\ \bibinfo {note}
  {arXiv:2405.04742}\BibitemShut {NoStop}%
\bibitem [{\citenamefont {Lewis-Swan}\ \emph
  {et~al.}(2019{\natexlab{c}})\citenamefont {Lewis-Swan}, \citenamefont
  {Safavi-Naini}, \citenamefont {Bollinger},\ and\ \citenamefont
  {Rey}}]{Le+Re19}%
  \BibitemOpen
  \bibfield  {author} {\bibinfo {author} {\bibfnamefont {R.~J.}\ \bibnamefont
  {Lewis-Swan}}, \bibinfo {author} {\bibfnamefont {A.}~\bibnamefont
  {Safavi-Naini}}, \bibinfo {author} {\bibfnamefont {J.~J.}\ \bibnamefont
  {Bollinger}},\ and\ \bibinfo {author} {\bibfnamefont {A.~M.}\ \bibnamefont
  {Rey}},\ }\href {https://doi.org/10.1038/s41467-019-09436-y} {\bibfield
  {journal} {\bibinfo  {journal} {Nat. Commun.}\ }\textbf {\bibinfo {volume}
  {10}},\ \bibinfo {pages} {1581} (\bibinfo {year}
  {2019}{\natexlab{c}})}\BibitemShut {NoStop}%
\bibitem [{\citenamefont {Zhou}\ and\ \citenamefont
  {Swingle}(2023{\natexlab{b}})}]{ZhSw23}%
  \BibitemOpen
  \bibfield  {author} {\bibinfo {author} {\bibfnamefont {T.}~\bibnamefont
  {Zhou}}\ and\ \bibinfo {author} {\bibfnamefont {B.}~\bibnamefont {Swingle}},\
  }\href {https://doi.org/10.1038/s41467-023-39065-5} {\bibfield  {journal}
  {\bibinfo  {journal} {Nature communications}\ }\textbf {\bibinfo {volume}
  {14}},\ \bibinfo {pages} {3411} (\bibinfo {year}
  {2023}{\natexlab{b}})}\BibitemShut {NoStop}%
\bibitem [{\citenamefont {Schuster}\ and\ \citenamefont
  {Yao}(2023)}]{SchYao23}%
  \BibitemOpen
  \bibfield  {author} {\bibinfo {author} {\bibfnamefont {T.}~\bibnamefont
  {Schuster}}\ and\ \bibinfo {author} {\bibfnamefont {N.~Y.}\ \bibnamefont
  {Yao}},\ }\href {https://doi.org/10.1103/physrevlett.131.160402} {\bibfield
  {journal} {\bibinfo  {journal} {Physical Review Letters}\ }\textbf {\bibinfo
  {volume} {131}},\ \bibinfo {pages} {160402} (\bibinfo {year}
  {2023})}\BibitemShut {NoStop}%
\bibitem [{\citenamefont {Braum{\"u}ller}\ \emph {et~al.}(2022)\citenamefont
  {Braum{\"u}ller}, \citenamefont {Karamlou}, \citenamefont {Yanay},
  \citenamefont {Kannan}, \citenamefont {Kim}, \citenamefont {Kjaergaard},
  \citenamefont {Melville}, \citenamefont {Niedzielski}, \citenamefont {Sung},
  \citenamefont {Veps{\"a}l{\"a}inen} \emph {et~al.}}]{Bra+Ol22}%
  \BibitemOpen
  \bibfield  {author} {\bibinfo {author} {\bibfnamefont {J.}~\bibnamefont
  {Braum{\"u}ller}}, \bibinfo {author} {\bibfnamefont {A.~H.}\ \bibnamefont
  {Karamlou}}, \bibinfo {author} {\bibfnamefont {Y.}~\bibnamefont {Yanay}},
  \bibinfo {author} {\bibfnamefont {B.}~\bibnamefont {Kannan}}, \bibinfo
  {author} {\bibfnamefont {D.}~\bibnamefont {Kim}}, \bibinfo {author}
  {\bibfnamefont {M.}~\bibnamefont {Kjaergaard}}, \bibinfo {author}
  {\bibfnamefont {A.}~\bibnamefont {Melville}}, \bibinfo {author}
  {\bibfnamefont {B.~M.}\ \bibnamefont {Niedzielski}}, \bibinfo {author}
  {\bibfnamefont {Y.}~\bibnamefont {Sung}}, \bibinfo {author} {\bibfnamefont
  {A.}~\bibnamefont {Veps{\"a}l{\"a}inen}}, \emph {et~al.},\ }\href
  {https://doi.org/10.1038/s41567-021-01430-w} {\bibfield  {journal} {\bibinfo
  {journal} {Nat. Phys.}\ }\textbf {\bibinfo {volume} {18}},\ \bibinfo {pages}
  {172} (\bibinfo {year} {2022})}\BibitemShut {NoStop}%
\bibitem [{\citenamefont {García-Mata}\ \emph {et~al.}(2023)\citenamefont
  {García-Mata}, \citenamefont {Jalabert},\ and\ \citenamefont
  {Wisniacki}}]{Garcia-Mata:2023}%
  \BibitemOpen
  \bibfield  {author} {\bibinfo {author} {\bibfnamefont {I.}~\bibnamefont
  {García-Mata}}, \bibinfo {author} {\bibfnamefont {R.~A.}\ \bibnamefont
  {Jalabert}},\ and\ \bibinfo {author} {\bibfnamefont {D.~A.}\ \bibnamefont
  {Wisniacki}},\ }\href {https://doi.org/10.4249/scholarpedia.55237} {\bibfield
   {journal} {\bibinfo  {journal} {Scholarpedia}\ }\textbf {\bibinfo {volume}
  {18}},\ \bibinfo {pages} {55237} (\bibinfo {year} {2023})}\BibitemShut
  {NoStop}%
\bibitem [{\citenamefont {Polchinski}(2015)}]{Pol15}%
  \BibitemOpen
  \bibfield  {author} {\bibinfo {author} {\bibfnamefont {J.}~\bibnamefont
  {Polchinski}},\ }\bibfield  {journal} {\bibinfo  {journal} {arXiv preprint
  arXiv:1505.08108}\ }\href {https://doi.org/10.48550/arXiv.1505.08108}
  {10.48550/arXiv.1505.08108} (\bibinfo {year} {2015})\BibitemShut {NoStop}%
\bibitem [{\citenamefont {Rozenbaum}\ \emph {et~al.}(2017)\citenamefont
  {Rozenbaum}, \citenamefont {Ganeshan},\ and\ \citenamefont
  {Galitski}}]{RGGa17}%
  \BibitemOpen
  \bibfield  {author} {\bibinfo {author} {\bibfnamefont {E.~B.}\ \bibnamefont
  {Rozenbaum}}, \bibinfo {author} {\bibfnamefont {S.}~\bibnamefont
  {Ganeshan}},\ and\ \bibinfo {author} {\bibfnamefont {V.}~\bibnamefont
  {Galitski}},\ }\href {https://doi.org/10.1103/PhysRevLett.118.086801}
  {\bibfield  {journal} {\bibinfo  {journal} {Phys. Rev. Lett.}\ }\textbf
  {\bibinfo {volume} {118}},\ \bibinfo {pages} {086801} (\bibinfo {year}
  {2017})}\BibitemShut {NoStop}%
\bibitem [{\citenamefont {Garc{\'{\i}}a-Mata}\ \emph
  {et~al.}(2018)\citenamefont {Garc{\'{\i}}a-Mata}, \citenamefont {Saraceno},
  \citenamefont {Jalabert}, \citenamefont {Roncaglia},\ and\ \citenamefont
  {Wisniacki}}]{Gm+Wi18}%
  \BibitemOpen
  \bibfield  {author} {\bibinfo {author} {\bibfnamefont {I.}~\bibnamefont
  {Garc{\'{\i}}a-Mata}}, \bibinfo {author} {\bibfnamefont {M.}~\bibnamefont
  {Saraceno}}, \bibinfo {author} {\bibfnamefont {R.~A.}\ \bibnamefont
  {Jalabert}}, \bibinfo {author} {\bibfnamefont {A.~J.}\ \bibnamefont
  {Roncaglia}},\ and\ \bibinfo {author} {\bibfnamefont {D.~A.}\ \bibnamefont
  {Wisniacki}},\ }\bibfield  {journal} {\bibinfo  {journal} {Phys.Rev.Lett.}\
  }\textbf {\bibinfo {volume} {121}},\ \href
  {https://doi.org/10.1103/physrevlett.121.210601}
  {10.1103/physrevlett.121.210601} (\bibinfo {year} {2018})\BibitemShut
  {NoStop}%
\bibitem [{\citenamefont {Fortes}\ \emph {et~al.}(2019)\citenamefont {Fortes},
  \citenamefont {Garc\'{\i}a-Mata}, \citenamefont {Jalabert},\ and\
  \citenamefont {Wisniacki}}]{FGmJaWi19}%
  \BibitemOpen
  \bibfield  {author} {\bibinfo {author} {\bibfnamefont {E.~M.}\ \bibnamefont
  {Fortes}}, \bibinfo {author} {\bibfnamefont {I.}~\bibnamefont
  {Garc\'{\i}a-Mata}}, \bibinfo {author} {\bibfnamefont {R.~A.}\ \bibnamefont
  {Jalabert}},\ and\ \bibinfo {author} {\bibfnamefont {D.~A.}\ \bibnamefont
  {Wisniacki}},\ }\href {https://doi.org/10.1103/PhysRevE.100.042201}
  {\bibfield  {journal} {\bibinfo  {journal} {Phys. Rev. E}\ }\textbf {\bibinfo
  {volume} {100}},\ \bibinfo {pages} {042201} (\bibinfo {year}
  {2019})}\BibitemShut {NoStop}%
\bibitem [{\citenamefont {Lozano-Negro}\ \emph {et~al.}(2021)\citenamefont
  {Lozano-Negro}, \citenamefont {Zangara},\ and\ \citenamefont
  {Pastawski}}]{LoZgPa21}%
  \BibitemOpen
  \bibfield  {author} {\bibinfo {author} {\bibfnamefont {F.}~\bibnamefont
  {Lozano-Negro}}, \bibinfo {author} {\bibfnamefont {P.~R.}\ \bibnamefont
  {Zangara}},\ and\ \bibinfo {author} {\bibfnamefont {H.~M.}\ \bibnamefont
  {Pastawski}},\ }\href {https://doi.org/10.1016/j.chaos.2021.111175}
  {\bibfield  {journal} {\bibinfo  {journal} {Chaos, Solitons and Fractals}\
  }\textbf {\bibinfo {volume} {150}},\ \bibinfo {pages} {111175} (\bibinfo
  {year} {2021})}\BibitemShut {NoStop}%
\bibitem [{\citenamefont {Pastawski}\ \emph {et~al.}(2000)\citenamefont
  {Pastawski}, \citenamefont {Levstein}, \citenamefont {Usaj}, \citenamefont
  {Raya},\ and\ \citenamefont {Hirschinger}}]{Pa+00}%
  \BibitemOpen
  \bibfield  {author} {\bibinfo {author} {\bibfnamefont {H.~M.}\ \bibnamefont
  {Pastawski}}, \bibinfo {author} {\bibfnamefont {P.~R.}\ \bibnamefont
  {Levstein}}, \bibinfo {author} {\bibfnamefont {G.}~\bibnamefont {Usaj}},
  \bibinfo {author} {\bibfnamefont {J.}~\bibnamefont {Raya}},\ and\ \bibinfo
  {author} {\bibfnamefont {J.}~\bibnamefont {Hirschinger}},\ }\href
  {https://doi.org/10.1016/S0378-4371(00)00146-1} {\bibfield  {journal}
  {\bibinfo  {journal} {Physica A: Statistical Mechanics and its Applications}\
  }\textbf {\bibinfo {volume} {283}},\ \bibinfo {pages} {166} (\bibinfo {year}
  {2000})}\BibitemShut {NoStop}%
\bibitem [{\citenamefont {Garc{\'\i}a-Mata}\ \emph {et~al.}(2022)\citenamefont
  {Garc{\'\i}a-Mata}, \citenamefont {Jalabert},\ and\ \citenamefont
  {Wisniacki}}]{GaJaWi22}%
  \BibitemOpen
  \bibfield  {author} {\bibinfo {author} {\bibfnamefont {I.}~\bibnamefont
  {Garc{\'\i}a-Mata}}, \bibinfo {author} {\bibfnamefont {R.~A.}\ \bibnamefont
  {Jalabert}},\ and\ \bibinfo {author} {\bibfnamefont {D.~A.}\ \bibnamefont
  {Wisniacki}},\ }\href@noop {} {\bibfield  {journal} {\bibinfo  {journal}
  {arXiv preprint arXiv:2209.07965}\ } (\bibinfo {year} {2022})}\BibitemShut
  {NoStop}%
\bibitem [{\citenamefont {Slichter}(1990)}]{Sl90}%
  \BibitemOpen
  \bibfield  {author} {\bibinfo {author} {\bibfnamefont {C.~P.}\ \bibnamefont
  {Slichter}},\ }\href@noop {} {\emph {\bibinfo {title} {Principles of magnetic
  resonance}}}\ (\bibinfo  {publisher} {Springer-Verlag},\ \bibinfo {address}
  {Berlin; New York},\ \bibinfo {year} {1990})\BibitemShut {NoStop}%
\bibitem [{\citenamefont {Haeberlen}(1976)}]{Hb76}%
  \BibitemOpen
  \bibfield  {author} {\bibinfo {author} {\bibfnamefont {U.}~\bibnamefont
  {Haeberlen}},\ }\href@noop {} {\emph {\bibinfo {title} {High Resolution {NMR}
  in solids}}}\ (\bibinfo  {publisher} {Academic Press},\ \bibinfo {year}
  {1976})\BibitemShut {NoStop}%
\bibitem [{\citenamefont {Ernst}\ \emph {et~al.}(1987)\citenamefont {Ernst},
  \citenamefont {Bodenhausen},\ and\ \citenamefont {Wokaun}}]{Er87}%
  \BibitemOpen
  \bibfield  {author} {\bibinfo {author} {\bibfnamefont {R.~R.}\ \bibnamefont
  {Ernst}}, \bibinfo {author} {\bibfnamefont {G.}~\bibnamefont {Bodenhausen}},\
  and\ \bibinfo {author} {\bibfnamefont {A.}~\bibnamefont {Wokaun}},\
  }\href@noop {} {\emph {\bibinfo {title} {Principles of nuclear magnetic
  resonance in one and two dimensions}}}\ (\bibinfo  {publisher} {Oxford Univ.
  Press},\ \bibinfo {address} {Oxford},\ \bibinfo {year} {1987})\BibitemShut
  {NoStop}%
\bibitem [{\citenamefont {Munowitz}\ and\ \citenamefont
  {Pines}(1987)}]{MuPi87b}%
  \BibitemOpen
  \bibfield  {author} {\bibinfo {author} {\bibfnamefont {M.}~\bibnamefont
  {Munowitz}}\ and\ \bibinfo {author} {\bibfnamefont {A.}~\bibnamefont
  {Pines}},\ }\href@noop {} {\emph {\bibinfo {title} {Principles and
  applications of multiple-quantum NMR}}},\ edited by\ \bibinfo {editor}
  {\bibfnamefont {S.~A.~R.}\ \bibnamefont {I.~Prigogine}}\ (\bibinfo
  {publisher} {John Wiley and Sons, Inc.},\ \bibinfo {year} {1987})\BibitemShut
  {NoStop}%
\bibitem [{\citenamefont {Cory}\ \emph {et~al.}(1997)\citenamefont {Cory},
  \citenamefont {Fahmy},\ and\ \citenamefont {Havel}}]{CoFH97}%
  \BibitemOpen
  \bibfield  {author} {\bibinfo {author} {\bibfnamefont {D.~G.}\ \bibnamefont
  {Cory}}, \bibinfo {author} {\bibfnamefont {A.~F.}\ \bibnamefont {Fahmy}},\
  and\ \bibinfo {author} {\bibfnamefont {T.~F.}\ \bibnamefont {Havel}},\ }\href
  {https://doi.org/10.1073/pnas.94.5.1634} {\bibfield  {journal} {\bibinfo
  {journal} {PNAS}\ }\textbf {\bibinfo {volume} {94}},\ \bibinfo {pages} {1634}
  (\bibinfo {year} {1997})}\BibitemShut {NoStop}%
\bibitem [{\citenamefont {Cory}\ \emph {et~al.}(2000)\citenamefont {Cory},
  \citenamefont {Laflamme}, \citenamefont {Knill}, \citenamefont {Viola},
  \citenamefont {Havel}, \citenamefont {Boulant}, \citenamefont {Boutis},
  \citenamefont {Fortunato}, \citenamefont {Lloyd}, \citenamefont {Martinez},
  \citenamefont {Negrevergne}, \citenamefont {Pravia}, \citenamefont {Sharf},
  \citenamefont {Teklemariam}, \citenamefont {Weinstein},\ and\ \citenamefont
  {Zurek}}]{Co+00}%
  \BibitemOpen
  \bibfield  {author} {\bibinfo {author} {\bibfnamefont {D.~G.}\ \bibnamefont
  {Cory}}, \bibinfo {author} {\bibfnamefont {R.}~\bibnamefont {Laflamme}},
  \bibinfo {author} {\bibfnamefont {E.}~\bibnamefont {Knill}}, \bibinfo
  {author} {\bibfnamefont {L.}~\bibnamefont {Viola}}, \bibinfo {author}
  {\bibfnamefont {T.~F.}\ \bibnamefont {Havel}}, \bibinfo {author}
  {\bibfnamefont {N.}~\bibnamefont {Boulant}}, \bibinfo {author} {\bibfnamefont
  {G.}~\bibnamefont {Boutis}}, \bibinfo {author} {\bibfnamefont
  {E.}~\bibnamefont {Fortunato}}, \bibinfo {author} {\bibfnamefont
  {S.}~\bibnamefont {Lloyd}}, \bibinfo {author} {\bibfnamefont
  {R.}~\bibnamefont {Martinez}}, \bibinfo {author} {\bibfnamefont
  {C.}~\bibnamefont {Negrevergne}}, \bibinfo {author} {\bibfnamefont
  {M.}~\bibnamefont {Pravia}}, \bibinfo {author} {\bibfnamefont
  {Y.}~\bibnamefont {Sharf}}, \bibinfo {author} {\bibfnamefont
  {G.}~\bibnamefont {Teklemariam}}, \bibinfo {author} {\bibfnamefont {Y.~S.}\
  \bibnamefont {Weinstein}},\ and\ \bibinfo {author} {\bibfnamefont {W.~H.}\
  \bibnamefont {Zurek}},\ }\href
  {https://doi.org/10.1002/1521-3978(200009)48:9/11<875::AID-PROP875>3.0.CO;2-V}
  {\bibfield  {journal} {\bibinfo  {journal} {Fortschritte der Physik}\
  }\textbf {\bibinfo {volume} {48}},\ \bibinfo {pages} {875} (\bibinfo {year}
  {2000})}\BibitemShut {NoStop}%
\bibitem [{\citenamefont {Munowitz}(1987)}]{Munowitz}%
  \BibitemOpen
  \bibfield  {author} {\bibinfo {author} {\bibfnamefont {M.}~\bibnamefont
  {Munowitz}},\ }\href@noop {} {\emph {\bibinfo {title} {Coherence and NMR}}}\
  (\bibinfo  {publisher} {John Wiley and Sons, Inc.},\ \bibinfo {year}
  {1987})\BibitemShut {NoStop}%
\bibitem [{\citenamefont {Khitrin}(1997)}]{Kh97}%
  \BibitemOpen
  \bibfield  {author} {\bibinfo {author} {\bibfnamefont {A.~K.}\ \bibnamefont
  {Khitrin}},\ }\href {https://doi.org/10.1016/s0009-2614(97)00661-1}
  {\bibfield  {journal} {\bibinfo  {journal} {Chem. Phys. Lett.}\ }\textbf
  {\bibinfo {volume} {274}},\ \bibinfo {pages} {217} (\bibinfo {year}
  {1997})}\BibitemShut {NoStop}%
\bibitem [{\citenamefont {Baum}\ and\ \citenamefont {Pines}(1986)}]{BPi86}%
  \BibitemOpen
  \bibfield  {author} {\bibinfo {author} {\bibfnamefont {J.}~\bibnamefont
  {Baum}}\ and\ \bibinfo {author} {\bibfnamefont {A.}~\bibnamefont {Pines}},\
  }\href {https://doi.org/10.1021/ja00284a001} {\bibfield  {journal} {\bibinfo
  {journal} {J. Am. Chem. Soc.}\ }\textbf {\bibinfo {volume} {108}},\ \bibinfo
  {pages} {7447} (\bibinfo {year} {1986})}\BibitemShut {NoStop}%
\bibitem [{\citenamefont {Suzuki}(1976)}]{Su76}%
  \BibitemOpen
  \bibfield  {author} {\bibinfo {author} {\bibfnamefont {M.}~\bibnamefont
  {Suzuki}},\ }\href@noop {} {\bibfield  {journal} {\bibinfo  {journal}
  {Communications in Mathematical Physics}\ }\textbf {\bibinfo {volume} {51}},\
  \bibinfo {pages} {183} (\bibinfo {year} {1976})}\BibitemShut {NoStop}%
\bibitem [{\citenamefont {De~Raedt}\ and\ \citenamefont
  {De~Raedt}(1983)}]{DRae83}%
  \BibitemOpen
  \bibfield  {author} {\bibinfo {author} {\bibfnamefont {H.}~\bibnamefont
  {De~Raedt}}\ and\ \bibinfo {author} {\bibfnamefont {B.}~\bibnamefont
  {De~Raedt}},\ }\href {https://doi.org/10.1103/PhysRevA.28.3575} {\bibfield
  {journal} {\bibinfo  {journal} {Phys. Rev. A}\ }\textbf {\bibinfo {volume}
  {28}},\ \bibinfo {pages} {3575} (\bibinfo {year} {1983})}\BibitemShut
  {NoStop}%
\bibitem [{\citenamefont {Raedt}\ and\ \citenamefont
  {Michielsen}(2006)}]{DRaeM04}%
  \BibitemOpen
  \bibfield  {author} {\bibinfo {author} {\bibfnamefont {H.}~\bibnamefont
  {Raedt}}\ and\ \bibinfo {author} {\bibfnamefont {K.}~\bibnamefont
  {Michielsen}},\ }\bibinfo {title} {Computational methods for simulating
  quantum computers},\ in\ \href@noop {} {\emph {\bibinfo {booktitle} {Host
  Publication}}}\ (\bibinfo {year} {2006})\ \bibinfo {note} {arXiv preprint
  quant-ph/0406210}\BibitemShut {NoStop}%
\bibitem [{\citenamefont {{\'{A}}lvarez}\ \emph {et~al.}(2008)\citenamefont
  {{\'{A}}lvarez}, \citenamefont {Danieli}, \citenamefont {Levstein},\ and\
  \citenamefont {Pastawski}}]{ADLP08}%
  \BibitemOpen
  \bibfield  {author} {\bibinfo {author} {\bibfnamefont {G.~A.}\ \bibnamefont
  {{\'{A}}lvarez}}, \bibinfo {author} {\bibfnamefont {E.~P.}\ \bibnamefont
  {Danieli}}, \bibinfo {author} {\bibfnamefont {P.~R.}\ \bibnamefont
  {Levstein}},\ and\ \bibinfo {author} {\bibfnamefont {H.~M.}\ \bibnamefont
  {Pastawski}},\ }\href {https://doi.org/10.1103/physrevlett.101.120503}
  {\bibfield  {journal} {\bibinfo  {journal} {Phys. Rev. Lett.}\ }\textbf
  {\bibinfo {volume} {101}},\ \bibinfo {pages} {120503} (\bibinfo {year}
  {2008})}\BibitemShut {NoStop}%
\bibitem [{\citenamefont {Dente}\ \emph {et~al.}(2013)\citenamefont {Dente},
  \citenamefont {Bederi{\'a}n}, \citenamefont {Zangara},\ and\ \citenamefont
  {Pastawski}}]{DBZgPa13}%
  \BibitemOpen
  \bibfield  {author} {\bibinfo {author} {\bibfnamefont {A.~D.}\ \bibnamefont
  {Dente}}, \bibinfo {author} {\bibfnamefont {C.~S.}\ \bibnamefont
  {Bederi{\'a}n}}, \bibinfo {author} {\bibfnamefont {P.~R.}\ \bibnamefont
  {Zangara}},\ and\ \bibinfo {author} {\bibfnamefont {H.~M.}\ \bibnamefont
  {Pastawski}},\ }\href@noop {} {\bibfield  {journal} {\bibinfo  {journal}
  {arXiv preprint arXiv:1305.0036}\ } (\bibinfo {year} {2013})}\BibitemShut
  {NoStop}%
\bibitem [{\citenamefont {Fel'dman}\ and\ \citenamefont
  {Lacelle}(1997)}]{Feld97}%
  \BibitemOpen
  \bibfield  {author} {\bibinfo {author} {\bibfnamefont {E.~B.}\ \bibnamefont
  {Fel'dman}}\ and\ \bibinfo {author} {\bibfnamefont {S.}~\bibnamefont
  {Lacelle}},\ }\href {https://doi.org/https://doi.org/10.1063/1.474949}
  {\bibfield  {journal} {\bibinfo  {journal} {J. Chem. Phys.}\ }\textbf
  {\bibinfo {volume} {107}},\ \bibinfo {pages} {7067} (\bibinfo {year}
  {1997})}\BibitemShut {NoStop}%
\bibitem [{\citenamefont {Rufeil-Fiori}\ \emph {et~al.}(2009)\citenamefont
  {Rufeil-Fiori}, \citenamefont {S\'{a}nchez}, \citenamefont {Oliva},
  \citenamefont {Pastawski},\ and\ \citenamefont {Levstein}}]{RfSaPa09}%
  \BibitemOpen
  \bibfield  {author} {\bibinfo {author} {\bibfnamefont {E.}~\bibnamefont
  {Rufeil-Fiori}}, \bibinfo {author} {\bibfnamefont {C.~M.}\ \bibnamefont
  {S\'{a}nchez}}, \bibinfo {author} {\bibfnamefont {F.~Y.}\ \bibnamefont
  {Oliva}}, \bibinfo {author} {\bibfnamefont {H.~M.}\ \bibnamefont
  {Pastawski}},\ and\ \bibinfo {author} {\bibfnamefont {P.~R.}\ \bibnamefont
  {Levstein}},\ }\href {https://doi.org/10.1103/PhysRevA.79.032324} {\bibfield
  {journal} {\bibinfo  {journal} {Phys. Rev. A}\ }\textbf {\bibinfo {volume}
  {79}},\ \bibinfo {pages} {032324} (\bibinfo {year} {2009})}\BibitemShut
  {NoStop}%
\bibitem [{\citenamefont {Cappellaro}\ \emph
  {et~al.}(2007{\natexlab{a}})\citenamefont {Cappellaro}, \citenamefont
  {Ramanathan},\ and\ \citenamefont {Cory}}]{CaRaCo07}%
  \BibitemOpen
  \bibfield  {author} {\bibinfo {author} {\bibfnamefont {P.}~\bibnamefont
  {Cappellaro}}, \bibinfo {author} {\bibfnamefont {C.}~\bibnamefont
  {Ramanathan}},\ and\ \bibinfo {author} {\bibfnamefont {D.~G.}\ \bibnamefont
  {Cory}},\ }\href {https://doi.org/10.1103/PhysRevLett.99.250506} {\bibfield
  {journal} {\bibinfo  {journal} {Phys. Rev. Lett.}\ }\textbf {\bibinfo
  {volume} {99}},\ \bibinfo {pages} {250506} (\bibinfo {year}
  {2007}{\natexlab{a}})}\BibitemShut {NoStop}%
\bibitem [{\citenamefont {Cappellaro}\ \emph
  {et~al.}(2007{\natexlab{b}})\citenamefont {Cappellaro}, \citenamefont
  {Ramanathan},\ and\ \citenamefont {Cory}}]{CaRaCo07b}%
  \BibitemOpen
  \bibfield  {author} {\bibinfo {author} {\bibfnamefont {P.}~\bibnamefont
  {Cappellaro}}, \bibinfo {author} {\bibfnamefont {C.}~\bibnamefont
  {Ramanathan}},\ and\ \bibinfo {author} {\bibfnamefont {D.~G.}\ \bibnamefont
  {Cory}},\ }\href {https://doi.org/10.1103/PhysRevA.76.032317} {\bibfield
  {journal} {\bibinfo  {journal} {Phys. Rev. A}\ }\textbf {\bibinfo {volume}
  {76}},\ \bibinfo {pages} {032317} (\bibinfo {year}
  {2007}{\natexlab{b}})}\BibitemShut {NoStop}%
\bibitem [{\citenamefont {Zhou}\ \emph {et~al.}(2020)\citenamefont {Zhou},
  \citenamefont {Xu}, \citenamefont {Chen}, \citenamefont {Guo},\ and\
  \citenamefont {Swingle}}]{Zh+Sw20}%
  \BibitemOpen
  \bibfield  {author} {\bibinfo {author} {\bibfnamefont {T.}~\bibnamefont
  {Zhou}}, \bibinfo {author} {\bibfnamefont {S.}~\bibnamefont {Xu}}, \bibinfo
  {author} {\bibfnamefont {X.}~\bibnamefont {Chen}}, \bibinfo {author}
  {\bibfnamefont {A.}~\bibnamefont {Guo}},\ and\ \bibinfo {author}
  {\bibfnamefont {B.}~\bibnamefont {Swingle}},\ }\href
  {https://doi.org/10.1103/PhysRevLett.124.180601} {\bibfield  {journal}
  {\bibinfo  {journal} {Phys. Rev. Lett.}\ }\textbf {\bibinfo {volume} {124}},\
  \bibinfo {pages} {180601} (\bibinfo {year} {2020})}\BibitemShut {NoStop}%
\bibitem [{\citenamefont {Hallatschek}\ and\ \citenamefont
  {Fisher}(2014)}]{HaFi14}%
  \BibitemOpen
  \bibfield  {author} {\bibinfo {author} {\bibfnamefont {O.}~\bibnamefont
  {Hallatschek}}\ and\ \bibinfo {author} {\bibfnamefont {D.~S.}\ \bibnamefont
  {Fisher}},\ }\href {https://doi.org/10.1073/pnas.1404663111} {\bibfield
  {journal} {\bibinfo  {journal} {Proceedings of the National Academy of
  Sciences}\ }\textbf {\bibinfo {volume} {111}},\ \bibinfo {pages} {E4911}
  (\bibinfo {year} {2014})}\BibitemShut {NoStop}%
\bibitem [{\citenamefont {Chatterjee}\ and\ \citenamefont
  {S.~Dey}(2016)}]{ShPar16}%
  \BibitemOpen
  \bibfield  {author} {\bibinfo {author} {\bibfnamefont {S.}~\bibnamefont
  {Chatterjee}}\ and\ \bibinfo {author} {\bibfnamefont {P.}~\bibnamefont
  {S.~Dey}},\ }\href {https://doi.org/10.1002/cpa.21571} {\bibfield  {journal}
  {\bibinfo  {journal} {Communications on Pure and Applied Mathematics}\
  }\textbf {\bibinfo {volume} {69}},\ \bibinfo {pages} {203} (\bibinfo {year}
  {2016})}\BibitemShut {NoStop}%
\bibitem [{\citenamefont {Foss-Feig}\ \emph {et~al.}(2015)\citenamefont
  {Foss-Feig}, \citenamefont {Gong}, \citenamefont {Clark},\ and\ \citenamefont
  {Gorshkov}}]{Foss+Gor15}%
  \BibitemOpen
  \bibfield  {author} {\bibinfo {author} {\bibfnamefont {M.}~\bibnamefont
  {Foss-Feig}}, \bibinfo {author} {\bibfnamefont {Z.-X.}\ \bibnamefont {Gong}},
  \bibinfo {author} {\bibfnamefont {C.~W.}\ \bibnamefont {Clark}},\ and\
  \bibinfo {author} {\bibfnamefont {A.~V.}\ \bibnamefont {Gorshkov}},\ }\href
  {https://doi.org/10.1103/PhysRevLett.114.157201} {\bibfield  {journal}
  {\bibinfo  {journal} {Physical review letters}\ }\textbf {\bibinfo {volume}
  {114}},\ \bibinfo {pages} {157201} (\bibinfo {year} {2015})}\BibitemShut
  {NoStop}%
\bibitem [{\citenamefont {S\'anchez}\ \emph {et~al.}(2009)\citenamefont
  {S\'anchez}, \citenamefont {Levstein}, \citenamefont {Acosta},\ and\
  \citenamefont {Chattah}}]{SaLeACh09}%
  \BibitemOpen
  \bibfield  {author} {\bibinfo {author} {\bibfnamefont {C.~M.}\ \bibnamefont
  {S\'anchez}}, \bibinfo {author} {\bibfnamefont {P.~R.}\ \bibnamefont
  {Levstein}}, \bibinfo {author} {\bibfnamefont {R.~H.}\ \bibnamefont
  {Acosta}},\ and\ \bibinfo {author} {\bibfnamefont {A.~K.}\ \bibnamefont
  {Chattah}},\ }\href {https://doi.org/10.1103/PhysRevA.80.012328} {\bibfield
  {journal} {\bibinfo  {journal} {Phys. Rev. A}\ }\textbf {\bibinfo {volume}
  {80}},\ \bibinfo {pages} {012328} (\bibinfo {year} {2009})}\BibitemShut
  {NoStop}%
\bibitem [{\citenamefont {Alvarez}\ and\ \citenamefont
  {Suter}(2011)}]{AlSu11a}%
  \BibitemOpen
  \bibfield  {author} {\bibinfo {author} {\bibfnamefont {G.~A.}\ \bibnamefont
  {Alvarez}}\ and\ \bibinfo {author} {\bibfnamefont {D.}~\bibnamefont
  {Suter}},\ }\href {https://doi.org/10.1103/PhysRevA.84.012320} {\bibfield
  {journal} {\bibinfo  {journal} {Phys. Rev. A}\ }\textbf {\bibinfo {volume}
  {84}},\ \bibinfo {pages} {012320} (\bibinfo {year} {2011})}\BibitemShut
  {NoStop}%
\bibitem [{\citenamefont {Álvarez}\ \emph {et~al.}(2013)\citenamefont
  {Álvarez}, \citenamefont {Kaiser},\ and\ \citenamefont
  {Suter}}]{Alvarez2013}%
  \BibitemOpen
  \bibfield  {author} {\bibinfo {author} {\bibfnamefont {G.~A.}\ \bibnamefont
  {Álvarez}}, \bibinfo {author} {\bibfnamefont {R.}~\bibnamefont {Kaiser}},\
  and\ \bibinfo {author} {\bibfnamefont {D.}~\bibnamefont {Suter}},\ }\href
  {https://doi.org/10.1002/andp.201300096} {\bibfield  {journal} {\bibinfo
  {journal} {Ann. Phys. (Berlin)}\ }\textbf {\bibinfo {volume} {525}},\
  \bibinfo {pages} {833} (\bibinfo {year} {2013})}\BibitemShut {NoStop}%
\bibitem [{\citenamefont {Sánchez}\ \emph
  {et~al.}(2023{\natexlab{b}})\citenamefont {Sánchez}, \citenamefont
  {Pastawski},\ and\ \citenamefont {Chattah}}]{SANCHEZ2023100104}%
  \BibitemOpen
  \bibfield  {author} {\bibinfo {author} {\bibfnamefont {C.}~\bibnamefont
  {Sánchez}}, \bibinfo {author} {\bibfnamefont {H.}~\bibnamefont
  {Pastawski}},\ and\ \bibinfo {author} {\bibfnamefont {A.}~\bibnamefont
  {Chattah}},\ }\href
  {https://doi.org/https://doi.org/10.1016/j.jmro.2023.100104} {\bibfield
  {journal} {\bibinfo  {journal} {Journal of Magnetic Resonance Open}\ ,\
  \bibinfo {pages} {100104}} (\bibinfo {year}
  {2023}{\natexlab{b}})}\BibitemShut {NoStop}%
\end{thebibliography}%

\end{document}